\documentclass[reprint,amsmath,amssymb,aps,prb, superscriptaddress,twocolumn]{revtex4-1}
\usepackage{graphicx}
\usepackage[colorlinks=true,citecolor=blue,linkcolor=blue,urlcolor=blue]{hyperref}

\usepackage[utf8]{inputenc}
\usepackage{amstext}			
\usepackage{amssymb}			
\usepackage{amsmath}			
\usepackage{amsfonts}
\usepackage{graphicx}
\usepackage{subfigure}
\usepackage{wrapfig}
\usepackage{esint}
\usepackage{hyperref}
\usepackage{multirow}
\usepackage{printlen}
\usepackage{placeins}
\usepackage[english]{babel}

\DeclareMathOperator{\tr}{Tr}

\newcommand{\ttr}[1]{\tr{\left[#1\right]}}
\newcommand{\lint}{\int\limits}
\newcommand{\lsum}{\sum\limits}
\newcommand{\BH}{\mathcal{H}}

\newcommand{\BQ}{\mathcal{Q}}
\newcommand{\BC}{\mathcal{C}}
\newcommand{\BM}{\mathcal{M}}

\newcommand{\BL}{\mathcal{L}}

\newcommand*\chem[1]{\ensuremath{\mathrm{#1}}}
\newcommand{\pmark}[1]{{\bf (#1)}}
\newcommand{\mprime}{^\prime}
\newcommand{\tprime}{$^\prime$}

\begin{document}
  \title{Free induction decays in nuclear spin-1/2 lattices with small number of interacting neighbors: the cases of silicon and fluorapatite}
  \date{3 November, 2019}
  \author{Grigory A. Starkov}
  \email{grigory.starkov@skolkovotech.ru}
  \affiliation{Skolkovo Institute of Science and Technology,  Skolkovo Innovation Center, Nobel Street 3, Moscow 143026, Russia}
\affiliation{Lebedev Physical Institute of the Russian Academy of Sciences, Leninsky prospect 53,
Moscow 119991, Russia}
  \author{Boris V. Fine}
  \email{b.fine@skoltech.ru}
   \affiliation{Skolkovo Institute of Science and Technology,  Skolkovo Innovation Center, Nobel Street 3, Moscow 143026, Russia}
\affiliation{Institute for Theoretical Physics, University of Heidelberg, Philosophenweg 12, 69120 Heidelberg, Germany}

\begin{abstract}
Nuclear spin-1/2 lattices where each spin has a small effective number of interacting neighbors represent a particular challenge for first-principles calculations of free induction decays (FIDs) observed by nuclear magnetic resonance (NMR).   The challenge originates from the fact that these lattices are far from the limit where classical spin simulations perform well. Here we use the recently developed method of hybrid quantum-classical simulations to compute nuclear FIDs for $^{29}$Si-enriched silicon and fluorapatite. In these solids, small effective number of interacting neighbors is either due to the partition of the lattice into pairs of strongly coupled spins (silicon), or due to the partition into strongly coupled chains (fluorapatite). We find a very good overall agreement between the hybrid simulation results and the experiments. In addition, we introduce an extension of the hybrid method, which we call the method of ``coupled quantum clusters''. It is tested on $^{29}$Si-enriched silicon and found to exhibit excellent performance.


 \end{abstract}

\maketitle

\section{Introduction}


Nuclear  free induction decay (FID) measured by nuclear magnetic resonace (NMR)  is a quantity proportional to an infinite temperature time auto-correlation function of the nuclear total spin polarization\cite{Lowe-57,Abragam-61}. The Fourier transform of the FID gives the shape of NMR absorption peak \cite{Bloch-46-1,Lowe-57,Abragam-61}. 
The measurements of  FIDs can be used to extract microscopic information about solids such as the distances between nuclear spins or electronic spin susceptibility.  
Beyond NMR, the simulations of high-temperature spin dynamics belong to the broader field of dynamic thermalization.

First-principles calculation of NMR FID in solids is a long-standing problem, which is almost as old as the field of NMR itself \cite{Bloch-46-1, Bloch-46-2,  Lowe-57}. 
Quite a number of methods for the first-principles calculations were proposed in the past \cite{VanVleck-48,Lowe-57,Abragam-61,Tjon-66,Parker-73,Jensen-73,Engelsberg-75,Becker-76,Shakhmuratov-91,Lundin-92,Jensen-95,Lundin-96,Fine-97,Zhang-07,Savostyanov-14,Elsayed-15},
however, none of them is widely applied today.
This situation is, in part due to the non-perturbative character of the FID problem: there is no clear separation of time-scales, hence, there is no apparent small parameter for an approximate expansion.
As a result, the above-cited methods were, typically, based on uncontrolled approximations.
Another reason for the absence of a widespread adoption of a single method is that the FID approximation schemes were rarely tested beyond the case of the NMR benchmark material calcium fluoride ($\text{CaF}_2$)\cite{Engelsberg-74}.
As a consequence, the predictive performance of these schemes for a broader class of systems remained unclear.

Recently, we proposed\cite{Starkov-18}
a hybrid quantum-classical method based on simulating a large quantum spin lattice by
a small cluster of quantum spins coupled to an environment of interacting classical spins {\it via}  a correlation-preserving scheme. The unique feature of the method is that it affords an effective estimate of the uncertainty of its predictions by comparing the results of simulations for different sizes of the quantum cluster. This means that the reliability of the hybrid predictions can be assessed without comparing with an experiment or with a numerically exact quantum result.
 In \cite{Starkov-18}, we extensively tested the hybrid method on various model one- and two-dimensional lattices of spins-$1/2$ with nearest-neighbor interactions, as well as on the experimentally measured FIDs in $\text{CaF}_2$\cite{Engelsberg-74}.
In almost all the cases the observed performance was excellent, and when it was not, the above-mentioned uncertainty estimate indicated a discrepancy prior to the comparison with the reference data.

For spin lattices where each spin strongly interacts with a sufficiently large number of neighbors $n_\text{eff}$, purely classical simulations were found  to describe the FIDs quite accurately\cite{Elsayed-15}. An example here is  $\text{CaF}_2$.
In this case, the hybrid method generates results that exhibit only small deviations from the classical predictions. Hybrid calculations are still useful for large-$n_\text{eff}$ lattices, because the deviation between the classical and the hybrid results quantifies the predictive uncertainty of the both methods\cite{Starkov-18}.  
However, the true value of the hybrid method is in the simulations of three-dimensional spin-1/2 lattices with small $n_\text{eff}$. In such a setting, classical simulations are not expected to be quantitatively accurate, while  direct  purely quantum simulations are not feasible.

Two examples of small-$n_\text{eff}$  spin-1/2 systems are $^{29}\text{Si}$-enriched silicon, where, for certain orientations of external magnetic field, the lattice breaks into strongly interacting spin pairs,
and fluorapatite $\chem{Ca}_{10}(\chem{PO}_4)_6 \chem{F}_2$, where $^{19}$F nuclei are positioned in parallel chains with weak interchain coupling. In the present work, we test the hybrid method by comparing its predictions with the measured FIDs in $^{29}\text{Si}$-enriched silicon and  fluorapatite. In addition, we introduce an extension of the hybrid method, which we call the method of ``coupled quantum clusters''. The latter method is tested on $^{29}$Si-enriched silicon.



\section{General formulation}\label{hybrid_overview}

Let us consider a material with one type of magnetic nuclei and no disorder.

The FID experiment in solids measures the relaxation of the total spin magnetization transverse to
a strong static magnetic field $\mathbf{B}_0$.
In the Larmor rotating reference frame, the relaxation is described by the effective truncated Hamiltonian of the general form
\begin{equation}
    {\BH} = \lsum_{\alpha,i<j} J_{i,j}^\alpha {S}_i^{\alpha}{S}_j^{\alpha},\qquad \alpha\in\{x,y,z\},
    \label{Ham}
\end{equation}
where $S_i^\alpha$ is the operator of spin projection on axis $\alpha$ for the $i$-th lattice site
and the $z$-axis is chosen along the direction of $\mathbf{B}_0$.
The coupling constants $J^\alpha_{i,j}$ correspond to the magnetic dipolar interaction between nuclear spins averaged over the fast Larmor precession. They have form
\begin{equation}
 J^z_{i,j} = -2J^x_{i,j} = -2J^y_{i,j} = \cfrac{\gamma^2\hbar^2(1-3 \cos^2{\theta_{ij}})}{|{\bf r}_{ij}|^3}.
 \label{dipolar}
\end{equation}
Here, 
${\bf r}_{ij}$ is the vector connecting lattice sites $i$ and $j$, $\theta_{ij}$ is the angle between ${\bf r}_{ij}$ and $\mathbf{B}_0$, $\gamma$ is the gyromagnetic ratio of nuclear spins.

The effective number of interacting neighbours, which controls the applicability of the classical simulations, is defined as \cite{Elsayed-15}
\begin{equation}
n_{\text{eff}}  \equiv 
\frac{
\left[  
\sum_i \left( {J^x_{i,j}}^{\!\! 2}  +   {J^y_{i,j}}^{\!\! 2}   +   {J^z_{i,j}}^{\!\! 2}  \right) 
 \right]^2
 }
 {  
\sum_i  \left( {J^x_{i,j}}^{\!\! 2}  +   {J^y_{i,j}}^{\!\! 2}   +   {J^z_{i,j}}^{\!\! 2}    \right)^2
 }.
 \label{neff}
 \end{equation}

The signal measured in an FID experiment is proportional to the equilibrium infinite-temperature time auto-correlation function $C_x(t)$ of the total spin polarization $M_x(t)$ along transverse direction:
\begin{equation}
   C_x(t) = \left\langle M_x(t) M_x(0)\right\rangle  , 
   \label{corf1}
\end{equation}
where
\begin{equation}
   {M}_x = \sum_i{S}_i^x.
   \label{corf1_M}
\end{equation}
In the case of purely quantum dynamics, the notation  $\langle ... \rangle$ in Eq.(\ref{corf1}) is defined as:
\begin{equation}
 \left\langle M_x(t) M_x(0)\right\rangle = \cfrac1D \ttr{ M_x(t) M_x(0)},
 \label{MM}
\end{equation}
where $D$ is the dimensionality of the Hilbert space of the entire lattice.



\section{Hybrid Method}

The idea of the hybrid method is to approximate the dynamics of the fully quantum lattice by that of the hybrid one consisting of a cluster of quantum spins surrounded by an environment of classical spins\cite{Starkov-18,Starkov-19d}. We denote the set of all sites of the hybrid lattice as $\BL$. Among them, we choose the subset of lattice sites $\BQ\in\BL$ for the spins of the quantum cluster, the latter being described by a wave-function $|\psi\rangle$.
The spins on the rest of the lattice sites $\BC = \BL/\BQ$ are treated classically, i. e. they are described as a set of three-dimensional vectors $\{\mathbf{s}_m\}$. The entire hybrid lattice has periodic boundary conditions.

The evolution of the quantum and the classical parts of the system are determined by the quantum and the classical Hamiltonians $\BH_\BQ$ and $H_\BC$ respectively:
\begin{align}
   {\BH}_\BQ & = \lsum_{i<j,\alpha}^{i,j\in\BQ} J^\alpha_{i,j} {S}^\alpha_i {S}^\alpha_i - \lsum_{i\in\BQ} {\bf h}_i^{\BC\BQ} \cdot {\bf S}_i,\label{q_ham}\\
   H_\BC         & = \lsum_{m<n,\alpha}^{m,n\in\BC} J^\alpha_{m,n} s^\alpha_m s^\alpha_n -
                            \lsum_{m\in\BC} {\bf h}^{\BQ\BC}_m\cdot {\bf s}_m,\label{cl_ham}
\end{align}
where $S_i^\alpha$ are the operators of spin projections as in Eq.~\eqref{Ham},
${\bf s}_m\equiv (s_m^x,s_m^y,s_m^z)$ are vectors of length $\sqrt{S(S+1)}$ representing the classical spins. In this work, $S = 1/2$, hence  $\sqrt{S(S+1)} = \sqrt{3}/2$. Also, $\mathbf{h}^{\BC\BQ}_i$ and ${\bf h}_m^{\BQ\BC}$ are the effective magnetic fields coupling the quantum cluster and the classical environment to each other:
\begin{align}
   {\bf h}^{\BC\BQ}_i & = -\lsum_{n\in\mathcal{C}}\left(\begin{array}{c}
                                             J_{i,n}^x s_n^x \\
                                             J_{i,n}^y s_n^y \\
                                             J_{i,n}^z s_n^z \\
                                            \end{array}
\right),\label{hfield_cl}\\
   {\bf h}_m^{\BQ\BC} & = -\sqrt{D_\BQ+1} \ \lsum_{j\in\mathcal{Q}}\left(\begin{array}{c}
                        J_{m,j}^x \langle\psi|{S}_j^x|\psi\rangle \\
                        J_{m,j}^y \langle\psi|{S}_j^y|\psi\rangle \\
                        J_{m,j}^z \langle\psi|{S}_j^z|\psi\rangle \\
         \end{array}
\right),\label{backaction}
\end{align}
where $D_\BQ = 2^{N_\BQ}$ is the dimensionality of the Hilbert space of the quantum cluster consisting of $N_\BQ$ spins. The equilibrium noise of quantum expectation values $\langle\psi|S_i^\alpha|\psi\rangle$ is smaller than its classical counterpart $s_m^\alpha$ by the factor $1/\sqrt{D_\BQ+1}$ (for the explanation, see Refs.\cite{Starkov-18,Starkov-19d}). The above suppression is compensated by the factor $\sqrt{D_\BQ+1}$ in Eq.~\eqref{backaction}. This factor corrects the amplitude mismatch between ${\bf h}_m^{\BQ\BC}$ and ${\bf h}_m^{\BQ\BC}$ that would happen in its absence.

According to Hamiltonians~\eqref{q_ham} and~\eqref{cl_ham}, the equations of motion for the hybrid lattice take the form
\begin{align}
 |\dot{\psi}(t)\rangle & = -\cfrac{i}{\hbar}\BH_\BQ |\psi(t)\rangle\label{q_eq} \\
 \dot{\mathbf{s}}_m & = \mathbf{s}_m \times (\mathbf{h}_m^{\BC\BC} + \mathbf{h}_m^{\BQ\BC})\label{cl_eq},
\end{align}
where
\begin{equation}
 {\bf h}^{\BC\BC}_m = -\lsum^{n\in\mathcal{C}}_{n\neq m}\left(\begin{array}{c}
                                             J_{m,n}^x s_n^x \\
                                             J_{m,n}^y s_n^y \\
                                             J_{m,n}^z s_n^z \\
                                            \end{array}
\right).\label{hcl_field}
\end{equation}
The hybrid version of the total spin polarization is defined as
\begin{equation}
  \BM_x(t) = \sqrt{D_\BQ+1} \ \langle\psi(t)|\lsum_{i\in\BQ} {S}_i^x |\psi(t)\rangle
  + \lsum_{m\in\BC} s_m^x(t).
  \label{total_spin}
  \end{equation}
  
Our actual implementation of the hybrid method involves the following technical detail.
Due to the translational invariance of the original quantum system,
the quantum correlation function~\eqref{corf1} can be re-expressed as\cite{Starkov-18,Starkov-19d}
\begin{equation}
 C_x(t) = \cfrac{N_\BL}{N_{\BQ^\prime}}\langle M_x(t)  M^{\prime}_x\rangle,
 \label{autocorr_reduced}
\end{equation}
where $\BQ^\prime\in \BL$ is an arbitrary subset of lattice sites, $N_\BL$ is the total number of spins in the lattice, $N_{\BQ^\prime}$ is the number of spins in the subset $\BQ^\prime$, and
\begin{equation}
  M^{\prime}_x = \lsum_{i\in\BQ^\prime} S_i^x.
  \label{MQp}
\end{equation}

Once we transition to the hybrid dynamics, the presence of the quantum-classical border breakes the translational invariance of the lattice, thereby making different choices of $\BQ^\prime$ inequivalent.
In order to minimize the influence of the quantum-classical border,
we take $\BQ^\prime$ in the fully quantum definition (\ref{autocorr_reduced}) to consist of one or several central spins within the quantum cluster $\BQ$.

Finally, the exact quantum  correlation function \eqref{autocorr_reduced} is replaced by the one for the infinite-temperature equilibrium noise generated by the hybrid dynamics:
\begin{equation} 
  C_{x} (t) = \cfrac{N_\BL}{N_{\BQ^\prime}}\left[ \BM_x(t) \BM_x'(0) \right]_{i.c.},
  \label{hybrid_autocorr}
\end{equation}
where $\left[\dotsb\right]_{i.c.}$ denotes the averaging over initial conditions, $\BM_x(t)$ is given by Eq.(\ref{total_spin}) and
\begin{equation}
 \BM_x' (t) = \sqrt{D_\BQ+1}\cdot\langle\psi(t)|\lsum_{m\in\BQ'} {S}_m^x |\psi(t)\rangle.\label{central_spins}
\end{equation}

The infinite-temperature ensemble of initial conditions is generated through fully random choice of $|\psi(0)\rangle$ in the Hilbert space of the quantum cluster and fully random orientations of classical spins ${\bf s}_m^\alpha(0)$. The time evolutions of $|\psi(t)\rangle$ and ${\bf s}_m^\alpha(t)$ are computed using the 4th-order Runge-Kutta routines for direct time integration as in Refs.\cite{Elsayed-13,Elsayed-15,Starkov-18,Starkov-19d}. The numbers and the durations of the computational runs behind the plots presented below are given in the Supplementary Material\cite{supplement}.

\section{Method of coupled quantum clusters}
\label{clusters}

Classical spins in the hybrid method can be thought of as each representing the quantum-mechanical expectation values for a quantum cluster consisting of one spin $1/2$\cite{Starkov-18}. From such a perspective, the hybrid method partitions the original quantum lattice into a larger central quantum cluster $\BQ$ and one-spin clusters represented by classical spins.

 In this section, we formulate a generalization of the hybrid method, which we call the ``method of coupled quantum clusters''.  It is based on  the partitioning the original quantum lattice into quantum clusters of arbitrary sizes and then coupling these clusters using the quantum-mechanical expectation values of the relevant observables. This method is to be tested in Section~\ref{silicon_diamond} on the FID in $^{29}\text{Si}$-enriched silicon, where the lattice can be naturally divided into pairs of spins 1/2.

Let us partition the quantum lattice defined by Hamiltonian~(\ref{Ham}) into smaller clusters. A cluster labeled by index $\mu$  contains a set of sites $\BQ_{\mu}$. It is represented by wave function $|\psi_{\mu}\rangle$ belonging to the Hilbert space of dimension $D_{\BQ_{\mu}} $.  The Hamiltonian for cluster $\BQ_{\mu}$ is defined as:
\begin{equation}
   {\BH}_{\BQ_{\mu}}  = \lsum_{i<j,\alpha}^{i,j\in\BQ_{\mu}} J^\alpha_{i,j} {S}^\alpha_i {S}^\alpha_i - \lsum_{i\in\BQ_{\mu}} {\bf h}_i^{\BQ_{\mu}} \cdot {\bf S}_i,
 \label{Qmu_ham}
\end{equation}
where 
\begin{equation}
   {\bf h}_i^{\BQ_{\mu}}  = - \lsum_{\nu}^{\nu \neq \mu}  \sqrt{D_{\BQ_{\nu}}+1} \ \lsum_{j\in\BQ_{\nu}}\left(\begin{array}{c}
                        J_{m,j}^x \langle\psi_{\nu}|{S}_j^x|\psi_{\nu}\rangle \\
                        J_{m,j}^y \langle\psi_{\nu}|{S}_j^y|\psi_{\nu}\rangle \\
                        J_{m,j}^z \langle\psi_{\nu}|{S}_j^z|\psi_{\nu}\rangle \\
         \end{array}
\right).
\label{hQmu}
\end{equation}

The dynamics of each wave function $|\psi_{\mu}\rangle$ is governed by the Schr{\"o}dinger equation $|\dot{\psi_{\mu}}(t)\rangle  = -\cfrac{i}{\hbar}\BH_{\BQ_{\mu}} |\psi(t)\rangle$. This leads to a system of coupled differential equations for all clusters, which is to be solved by the method of direct numerical time integration.
 
The ``clustered'' version of the total spin polarization is defined as
\begin{equation}
  \BM_x(t) = \lsum_{\mu} \sqrt{D_{\BQ_{\mu}}+1} \ \langle\psi_{\mu}(t)|\lsum_{i\in\BQ_{\mu}} {S}_i^x |\psi_{\mu}(t)\rangle.
  \label{total_spin_cl}
  \end{equation}
As in the hybrid method,  one can choose within each cluster $\BQ_{\mu}$ a subset of sites $\BQ'_{\mu}$  maximally separated from cluster's boundary and define
\begin{equation}
  \BM'_x(t) = \lsum_{\mu} \sqrt{D_{\BQ_{\mu}}+1} \ \langle\psi_{\mu}(t)|\lsum_{i\in\BQ'_{\mu}} {S}_i^x |\psi_{\mu}(t)\rangle.
  \label{total_spin_cl_pr}
  \end{equation}
Finally, the expression for the correlation function of interest is still given by Eq.(\ref{hybrid_autocorr}), but now the  definitions (\ref{total_spin_cl}) and (\ref{total_spin_cl_pr})  should be substituted there. 

The computational advantage of the method of coupled quantum clusters in comparison with the hybrid method is that each simulation run produces more statistically independent contributions to $ \BM'_x(t)$, because it tracks simultaneously many quantum clusters. The disadvantage, obviously, is that the time evolution of many quantum clusters is more computationally expensive to calculate than that of classical spins. Here, however, interesting compromises can be explored with not too large quantum clusters. 

The method of coupled quantum clusters as defined in this section is qualitatively different from the correlated cluster expansions reviewed in Ref.\cite{Yang-17} both in terms of the character of the simulations and in terms of the agenda. Our method aims at describing strongly correlated dynamics of a dense spin system, while the correlated cluster expansion targets the decoherence of a central spin in an environment of a dilute spin bath. 
At the same time, as far as the coupling scheme between different quantum clusters is concerned, the present method has certain parallels with the cluster truncated Wigner approximation proposed in Ref.\cite{Wurtz-18}. 

\section{FID for $^{29}\text{Si}$-enriched crystalline silicon }
\label{silicon_diamond}

\subsection{Preliminary remarks}

Crystalline silicon has a diamond-type crystal structure --- same as regular diamonds made of carbon atoms.
Both silicon and carbon have stable isotops with nuclear spins $1/2$:  $^{29}$Si and $^{13}$C, respectively.
The natural abundance of these isotopes is quite low:  $4.7\%$ for $^{29}$Si and $1.1\%$ for $^{13}$C.
However, crystals enriched to almost $100\%$ content of these particular isotopes have been grown artificially.

The FIDs of $99\%$ $^\text{13}$C-enriched diamond were measured in the past by Lefmann {\it et al.}~\cite{Lefmann-94} and Schaumburg {\it et al.}~\cite{Schaumburg-95} , while the FIDs of $96.9\%$ $^{\text{29}}\text{Si}$-enriched silicon
were measured by Verhulst {\it et al.}~\cite{Verhulst-03}.
The FID shapes obtained in the both cases are supposed to coincide once the time axes are properly rescaled, and, indeed, the two experimentally measured FIDs reasonably agree with each other provided experimental uncertainties are taken into account (see the Supplementary Material\cite{supplement}).  These uncertainties are, however, noticeable, as manifested, in particular, by the asymmetry of the measured absorption curves (the Fourier transforms of the FIDs) and by  the discrepancy between the experimental and the theoretical values
of second moments $M_2 \equiv -C_x^{\prime\prime}(0)/C_x(0)$,  the latter being computed for the truncated magnetic dipolar interaction. 

In the present work, we chose to compare the hybrid method predictions with the $^{29}$Si FIDs measured by  Verhulst {\it et al.}\cite{Verhulst-03}.  The gyromagnetic ratio for $^\chem{29}\chem{Si}$ is $\gamma = -5319\text{ rad s}^{-1}\text{ Oe}^{-1}$. Our theoretical calculations are to be done for $100\%$  $^\chem{29}\chem{Si}$-enriched samples.

On the theoretical side, the Fourier transforms of the FIDs for the diamond lattice were calculated: by Schaumburg {\it et al.}~\cite{Schaumburg-95}, who used the exact calculation of a 5-spin problem supplemented by a Gaussian broadening of the resulting lineshapes;   by Jensen~\cite{Jensen-95} with the help of a continued fraction representation of the Laplace transform of the FID; and by Lundin and Zobov~\cite{Lundin-18}, who relied on 
the scheme introduced in the 1996 work of Lundin~\cite{Lundin-96},
which, in turn, was based on the hypothesis of the asymptotic similarity of correlations functions of various orders.

\subsection{Lattice structure of silicon}
\label{lattice_Si}

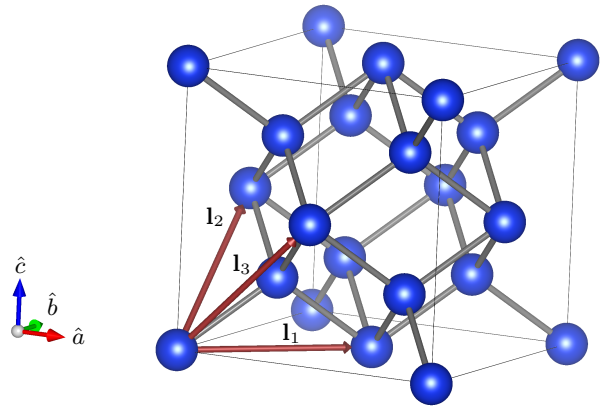
\begin{figure}\setlength{\unitlength}{0.1in}
 \begin{center}
 \begin{picture}(34, 22.8)(-0.5,0)
  \put(-3,0){\includegraphics[width = 4.7in]{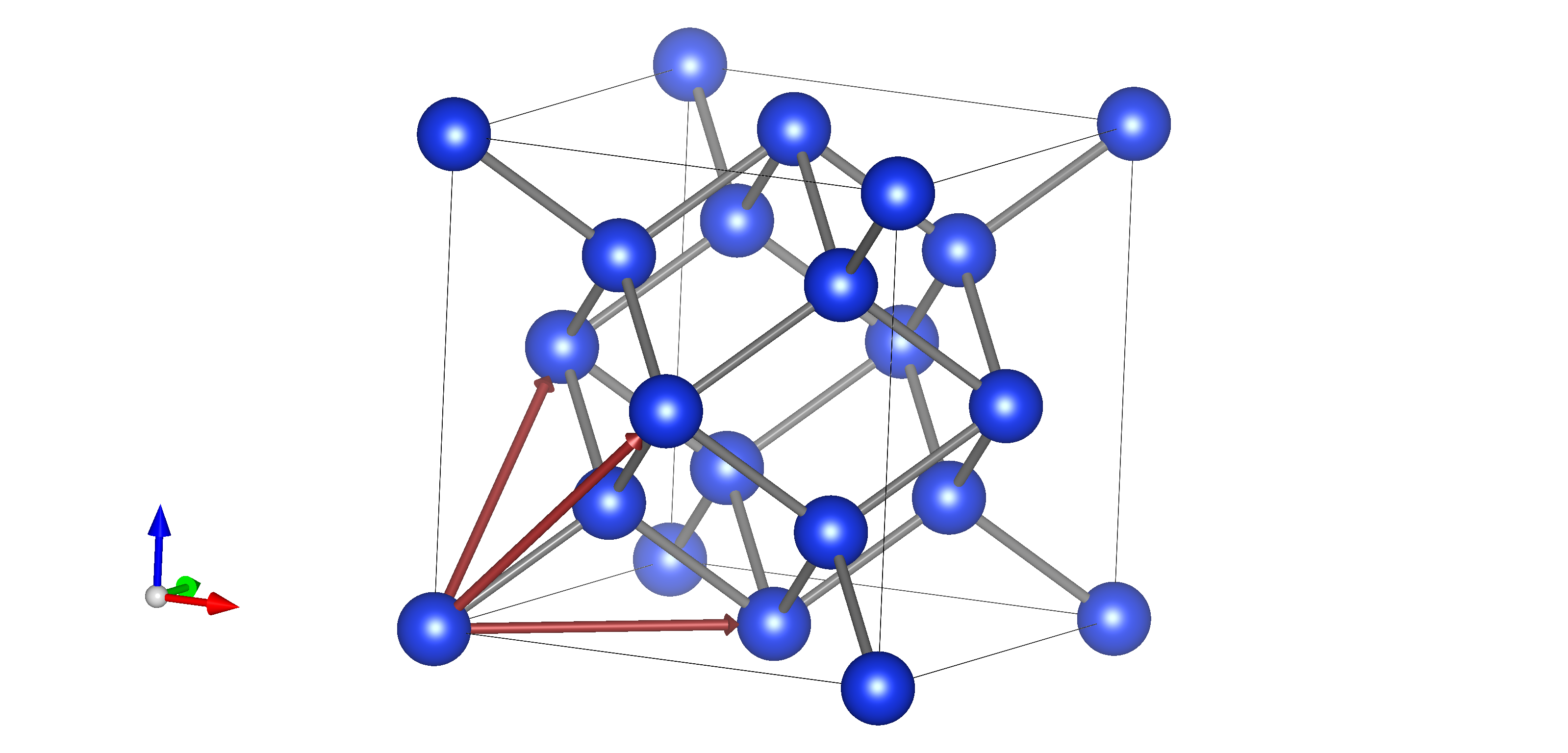}}
  \put(4.5,4){$\hat{a}$}
  \put(3.2,5.5){$\hat{b}$}
  \put(1.5,7.7){$\hat{c}$}
 
  \put(15.5,4.2){${\bf l}_1$}
  \put(11.5,10.3){${\bf l}_2$}
  \put(12.9,7.8){${\bf l}_3$}
 \end{picture}
  \caption{Diamond-type crystal structure of crystalline silicon.  The red arrows represent the primitive vectors of the lattice.}\label{diamond_cell}
  \end{center}
\end{figure}

The diamond-type crystal structure of silicon is presented on Fig.~\ref{diamond_cell}.
It is a face-centered cubic lattice with a two-site basis. The center of the unit cell is an inversion center of the lattice. As a consequence, two lattice sites of the unit cell are equivalent.
In terms of the orthonormal vectors $(\hat{a}, \hat{b}, \hat{c})$ shown in Fig.~\ref{diamond_cell},
the primitive vectors of the lattice are expressed as
\begin{equation}
 \mathbf{l}_1 = \frac{a_0}{2}(\hat{a}+\hat{b}), \quad \mathbf{l}_2 = \frac{a_0}{2}(\hat{b}+\hat{c}), \quad \mathbf{l}_3 = \frac{a_0}{2}(\hat{a}+\hat{c}),\label{simple_vectors}
\end{equation}
and two vectors of the basis are:
\begin{equation}
 \mathbf{v}_0 = \mathbf{0}, \quad \mathbf{v}_1 = \frac{a_0}{4}(\hat{a}+\hat{b}+\hat{c}),\label{simple_basis}
\end{equation}
where $a_0$ is the period of the fcc lattice (see also \cite{Ashcroft-76}).
For silicon diamond, $a_0 = 5.431$~\AA.

According to definition~(\ref{neff}), the effective numbers of interacting neighbors $n_\text{eff} $ for the external magnetic field $\mathbf{B}_0$ oriented along the [001], [011] and [111] crystal directions are, respectively,  27.4, 5.9 and 2.4.
As explained in the introduction, we are primarily interested in small $n_\text{eff} $, which are supposed to yield more rigorous tests of the hybrid method. Hence we primarily focus on the setting where
$\mathbf{B}_0$ parallel to the $[111]$ direction.  In this case, $n_\text{eff} $ is small, because each spin has one very strongly coupled neighbor along the $[111]$ direction [$\cos \theta_{ij} = \pm 1$ in Eq.(\ref{dipolar})]. The displacement vectors pointing at the three other nearest neighbors, while having the same length, are oriented with respect to [111] at angles corresponding to  $\cos \theta_{ij} = \mp1/3$, which makes the absolute value of the coupling constants (\ref{dipolar}) by a factor of three smaller than the largest one. 

In the above setting, the full quantum lattice can be naturally partitioned into pairs of  strongly coupled spins 1/2. The two spins within such a pair are displaced with respect to each other along the [111] direction. Each of them is  the strongest-coupled neighbor of the other one, which, in turn, implies that the interaction between different pairs is significantly smaller than the interaction within a pair. Given such a hierarchy, it is natural to expect that the method of coupled quantum clusters introduced in Section~\ref{clusters}  would be particularly efficient provided the strongly coupled spin pairs are chosen as quantum clusters $\BQ_{\mu}$ into which the full lattice is partitioned.

\subsection{Simulations vs. experiment for silicon}
\label{sim_exp_Si}

\begin{figure*}[t]
 \begin{center}
\includegraphics[width = 6.8in]{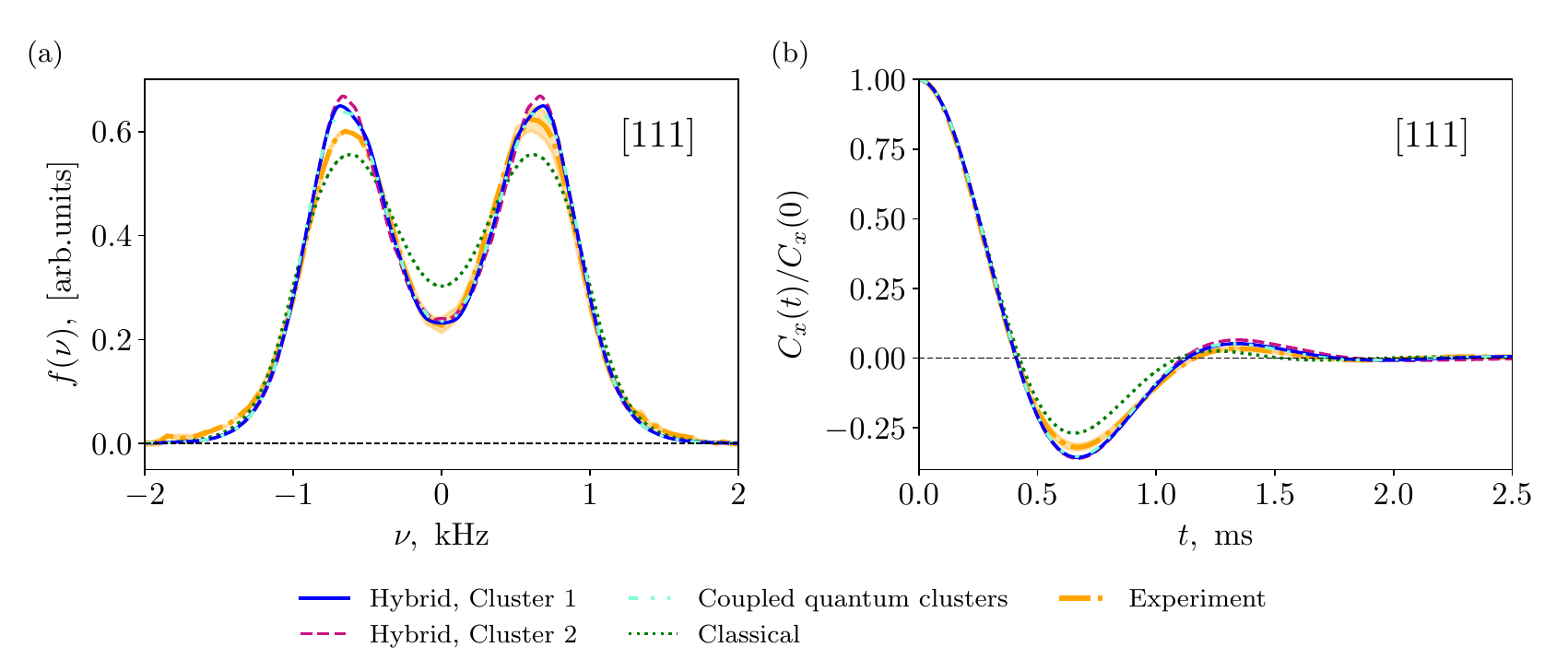}
  \caption{(a) Absorption  lineshape  and (b)  FID  in $^\text{29}$Si-enriched silicon for $\mathbf{B}_0$ along the  $[111]$ crystal direction:
  comparison of the results of simulations  with the experiment of Verhulst {\it et al.}\cite{Verhulst-03}. The simulations were done by the hybrid method and by the method of correlated quantum clusters.  The schemes of quantum clusters  1 and  2 used in the hybrid simulations are displayed in Fig.~\ref{cluster_schemes_111}. The coupled quantum cluster simulations were based on partitioning the full quantum lattice into pairs of strongly coupled spins 1/2 as described in the text. The experimental absorption lineshape is extracted directly from Ref.~\cite{Verhulst-03}, while the experimental FID is obtained by the Fourier transform of the absorption lineshape. The shaded area around the experimental lines is a measure of nonsystematic experimental error obtained as explained in \cite{supplement}.
 }
   \label{diamond_111}
 \end{center}
\end{figure*}

\begin{figure*}[t]\setlength{\unitlength}{0.1in}
\begin{center}
 \begin{picture}(60,20)
  \put(-2,0){\includegraphics[width=3.5in]{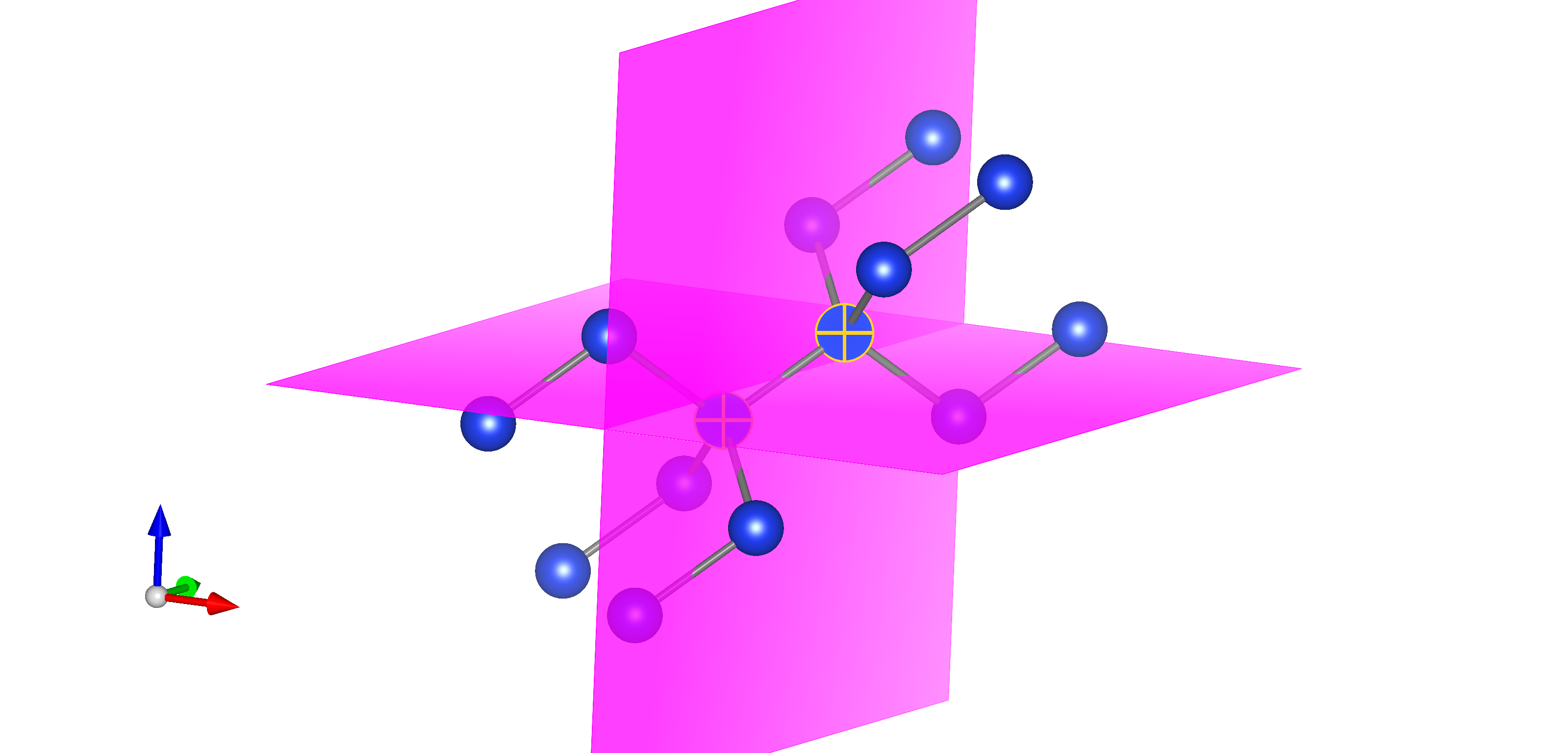}}
  \put(28,0){\includegraphics[width=3.5in]{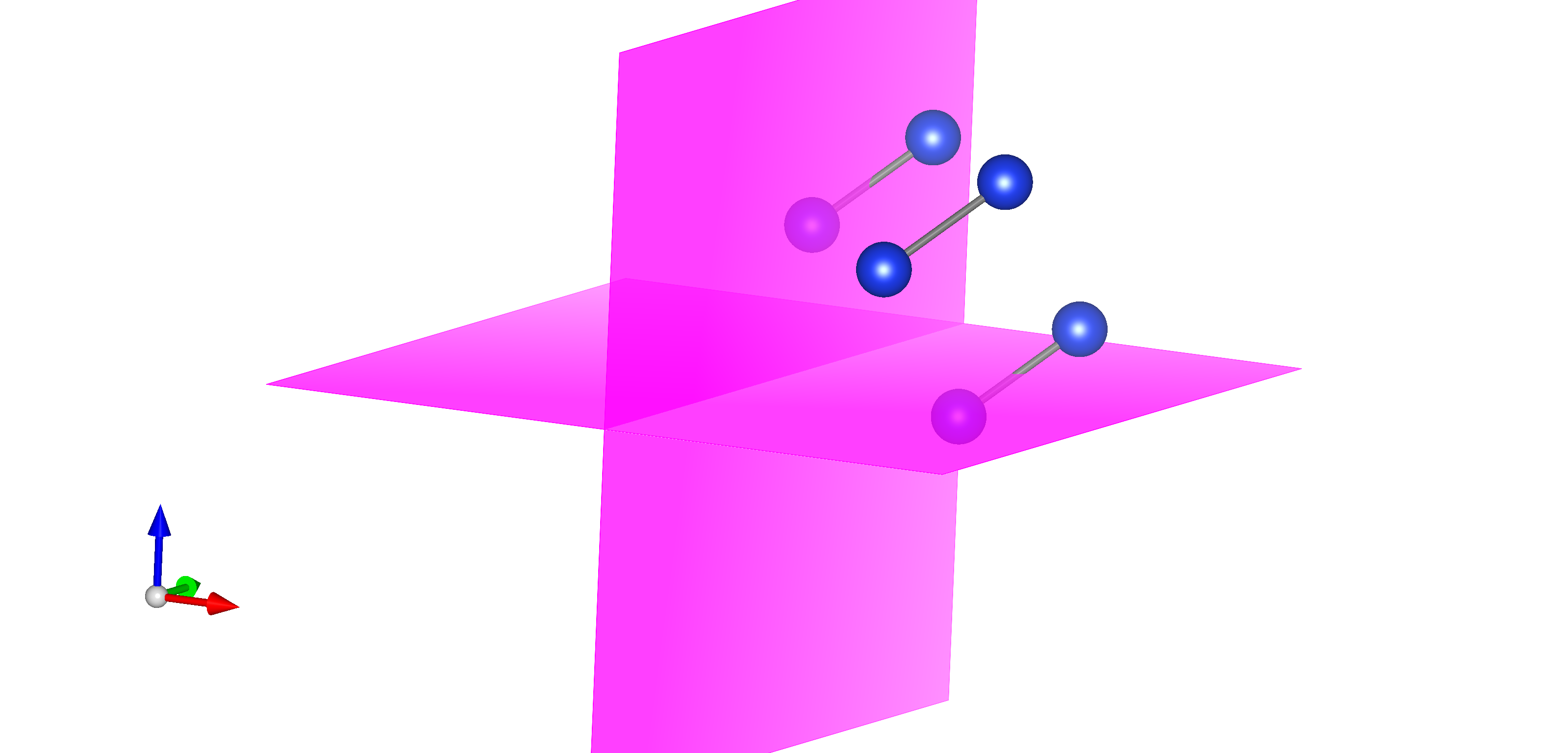}}
  \put(1,16){{\large (a)}}
  \put(31,16){{\large (b)}}
  
  \put(3.6,2.9){$\hat{a}$}
  \put(2.5,3.9){$\hat{b}$}
  \put(1.3, 5.8){$\hat{c}$}
  
  \put(33.6,2.9){$\hat{a}$}
  \put(32.5,3.9){$\hat{b}$}
  \put(31.3, 5.8){$\hat{c}$}

 \end{picture}
 \caption{Schemes of the quantum clusters $\BQ$  for the hybrid simulations of $^{\text{29}}\text{Si}$-enriched silicon for $\mathbf{B}_0\parallel [111]$ presented in Fig.~\ref{diamond_111}:  (a) and  (b) show cluster 1 and cluster 2 respectively.
 In (a), two sites belonging to the subset $\BQ'$ are marked with $+$. In (b), all sites belonging to $\BQ$ also belong to  $\BQ'$.
 }\label{cluster_schemes_111}
\end{center}
\end{figure*}
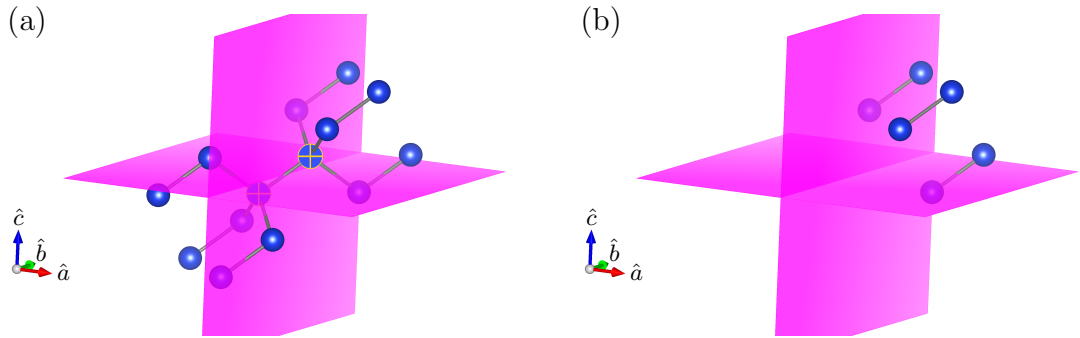

The results of our simulations for $\mathbf{B}_0$ along the $[111]$ direction both by the hybrid method and by the method of coupled quantum clusters  are presented and compared with experiment of Verhulst {\it et al.}\cite{Verhulst-03} in Fig.~\ref{diamond_111}.
Frame (a) displays the absorption lineshape $f(\nu)$, which is given by the Fourier transform of the FID:
\begin{equation}
 f(\nu) = \cfrac{2}{C_x(0)} \  \lint_0^{+\infty} dt \ C_x(t) \cos{2\pi\nu t}.
 \label{direct_transform}
\end{equation} 
Frame (b) displays the FID.

In order to estimate the accuracy of the hybrid simulations, we compare the hybrid results  for two different quantum clusters shown in Fig.~\ref{cluster_schemes_111}. The very small difference between the two results is a measure of the predictive uncertainty of the hybrid method. Detailed information about the simulations can be found in the Supplementary Material \cite{supplement}.

The simulations by the method of coupled quantum clusters  were performed by partitioning the full quantum lattice into clusters $\BQ_{\mu}$ consisting of pairs of strongly coupled spins 1/2 as described at the end of Section~\ref{lattice_Si}. We also chose $\BQ_{\mu}' = \BQ_{\mu}$. The plots for the coupled quantum clusters in Fig.~\ref{diamond_111} nearly coincide with the hybrid plots. At the same time, the convergence of the statistical averaging for  coupled quantum clusters method is significantly faster than that for the hybrid method.

As one can see in Fig.~\ref{diamond_111},  the agreement between both kinds of simulations and the experiment is very good, and, moreover, the small residual discrepancy might be due to experimental uncertainties or due to microscopic details not included into the model Hamiltonian. The former can be quantified through the ratio 1.33 of the experimental and the first-principles theoretical values of the second FID moments\cite{supplement}. The latter can be associated with crystal defects, paramagnetic impurities or less than $100\%$ abundance of $^\chem{29}\chem{Si}$.  As further illustrated in Fig.~\ref{diamond_111_theory} of Appendix~\ref{extra_si_sims},   the theoretical predictions of Jensen and of Lundin and Zobov appear to exhibit somewhat larger deviations from the experiment. 

It is worth remarking that the absorption lineshape in Fig.~\ref{diamond_111}(a) inherits its two-peak structure from the Pake doublet\cite{Pake-48} associated with an isolated pair of spins 1/2. Pake doublet  is sometimes viewed as an essentially quantum phenomenon originating from the dicreteness of quantum energy levels. Yet, even in this case, simulations of two classical spins  were shown\cite{Henner-16} to qualitatively reproduce the two-peak character of the absorption lineshape. This is an example of a rather subtle relation between classical and quantum dynamics: on the one hand, classical spin systems can be useful for practical calculations\cite{Jensen-73,Lundin-77,Tang-92,Elsayed-15} and also exhibit significant qualitative similarities with quantum ones as far as the long-time relaxation is concerned\cite{Fine-03,Fine-04,Fine-05,Morgan-08,Sorte-11,Meier-12} ; on the other hand, classical spin lattices are chaotic\cite{deWijn-12,deWijn-13}, while quantum lattices are not\cite{Fine-14} in the sense of the absence of the Lyapunov regime, even though they can imitate the Lyapunov regime over a limited time range\cite{Elsayed-15q}. 
In Fig.~\ref{diamond_111}, in order to highlight the difference between quantum and classical FIDs for small $n_\text{eff} $, we also include  the results of purely classical simulations of the kind done in Ref.\cite{Elsayed-15}.

Finally, we also performed hybrid  and classical simulations for  $\mathbf{B}_0$ parallel to the $[011]$ and $[001]$ crystal directions. The results are presented in Appendix~\ref{extra_si_sims}.

\section{$^{19}$F FID for fluorapatite}
\label{fluor_section}

\subsection{Preliminary discussion}

Fluorapatite $\chem{Ca}_{10}(\chem{PO}_4)_6 \chem{F}_2$ is a material often used to study spin dynamics of low-dimensional lattices \cite{Cappellaro-07a, Cappellaro-07, Zhang-09, Ramanathan-11, Kaur-13, Doronin-15, Bochkin-19}: Fluorine nuclei in fluorapatite   are arranged in parallel chains. For the orientation of external magnetic field parallel to the chains, the interactions between them are much smaller than the interaction within the chain.
As a result, $n_\text{eff}\approx2$, which means the fluorapatite lattice with coupling constants (\ref{dipolar}) can be viewed as a collection of weakly coupled spin chains.  In this section, we focus on $^{19}\text{F}$ FID in the above setting. We compute this FID and compare the result with  the measurements of  Engelsberg, Lowe and Carolan \cite{Engelsberg-73}.

In terms of contributions to nuclear magnetism, fluorapatite contains stable nuclear isotopes $^{19}\chem{F}$ and $^{31}\chem{P}$, which have spins 1/2 and natural abundances  $100\%$. We include both of them in the simulations. Their gyromagnetic ratios are $\gamma_F = 25166.2 \text{ rad s}^{-1}\text{ Oe}^{-1}$ and $\gamma_P = 10829.1 \text{ rad s}^{-1}\text{ Oe}^{-1}$ respectively. At the same time, magnetically active isotopes of calcium and oxygen have natural abundances less then $1\%$; hence, we neglect them.


\subsection{Lattice structure of fluorapatite}
\label{fluor_structure}

\begin{figure}\setlength{\unitlength}{0.1in}
 \begin{center}
 \begin{picture}(28, 18.5)
  \put(0,0){\includegraphics[width=2.8in]{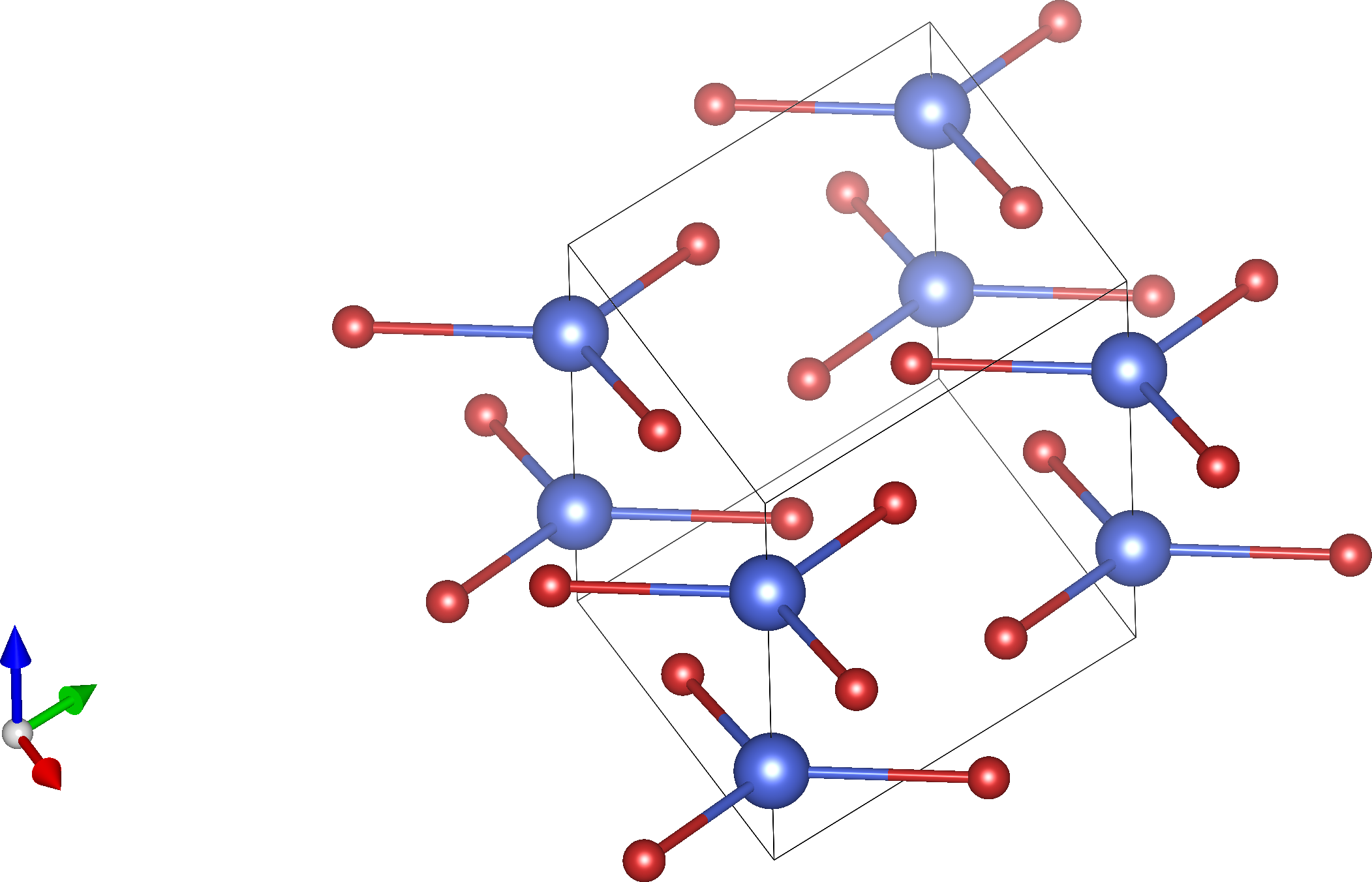}}
  
  \put(1.4,1.2){$\hat{a}$}
  \put(2.2, 4.0){$\hat{b}$}
  \put(0.0,5.7){$\hat{c}$}

  \end{picture}
  \caption{Scheme of a unit cell of fluorapatite. Only fluorine (blue) and phosphorous (red) atoms are shown.
 }\label{fluor_unit_cell}
 \end{center}
\end{figure}
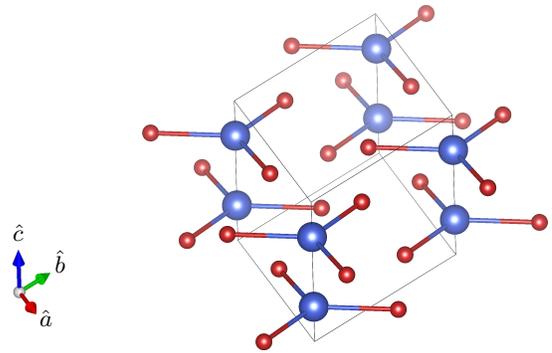

Fluorapatite has hexagonal crystal structure with the space group $P6_3/m$~\cite{Leroy-01}.
The lattice parameters are $a = b = 9.462$~\AA{} and $c = 6.849$~\AA. We denote the respective primitive vectors as $\mathbf{a}$, $\mathbf{b}$ and $\mathbf{c}$. The angle between $\mathbf{a}$ and $\mathbf{b}$ is $120^\circ$, and 
the $c$-axis is orthogonal to the hexagonal $ab$-plane. 
The basis cell of the sublattice of magnetically active nuclei contains two F nuclei at positions
\begin{equation}
 [0.0,0.0,0.25],\quad [0.0,0.0,0.75]
\end{equation}
and six P nuclei at positions
\begin{equation}
 \begin{array}{lll}
  [x,y,0.25], & [1-y, x-y, 0.25],\\ \relax
  [y-x,1-x,0.25], & [1-x, 1-y, 0.75],\\ \relax
   [y,y-x.0.75], & [x-y, x, 0.75],
 \end{array}
\end{equation}
where $x = 0.369$ and $y = 0.3985$. The coordinates are given in the basis of vectors $\mathbf{a}$, $\mathbf{b}$ and $\mathbf{c}$.
An illustration of the unit cell of fluorapatite is presented in Fig.~\ref{fluor_unit_cell}.
The positions of the $^{19}\chem{F}$ nuclei inside the basis cell are equivalent, since they are transformed into each other by  the discrete symmetry transformations of the lattice.
The positions of the $^{31}\chem{P}$ nuclei inside the basis cell are equivalent as well. 

The above mentioned strongly coupled chains of $^{19}\chem{F}$ nuclei  extend along the [001] direction.  The inter-chain distance betwenn nuclei is approximately $2.8$ times smaller then the intra-chain one. In the case where the external magnetic field $\mathbf{B}_0$ is parallel to the [001] direction, the largest value of intra-chain coupling is at least $21$ times smaller then the nearest-neighbour coupling within a chain. 

In comparison with CaF$_2$ and silicon, the simulations of $^{19}\text{F}$ FID in fluorapatite are complicated by the presence of ``unlike'' $^{31}\text{P}$ nuclei and by lattice disorder. Below we introduce technical modifications requred to accommodate these two aspects.

\subsection{Unlike spins}

Two nuclear spins with different gyromagnetic ratios are referred to in NMR literature as ``unlike spins''\cite{Abragam-61}. The truncated Hamiltonian averaged over fast precession of $^{19}\text{F}$ and $^{31}\text{P}$ nuclear spins, takes the form similar to Eqs.~\eqref{Ham} and~\eqref{dipolar}:
\begin{multline}
 \BH = \\ \* \lsum_{i<j,\alpha} {J}^\alpha_{i,j} S_i^\alpha S_j^\alpha
       + \lsum_{k<l,\alpha} \tilde{\tilde{{J}}}^\alpha_{k,l} I_i^\alpha I_j^\alpha
       + \lsum_{i,k,\alpha} \tilde{J}^\alpha_{i,k} S_i^\alpha I_k^\alpha.\label{truncated_dipolar_unlike}
\end{multline}
Here, $S_i^\alpha$ and $I_k^\alpha$ are the spin operators of $^{19}\text{F}$ and $^{31}\text{P}$ nuclei respectively. The coupling constants $J_{i,j}^\alpha$, $\tilde{\tilde{{J}}}^\alpha_{k,l}$ and $\tilde{J}^\alpha_{i,k}$ are
\begin{align}
 J_{i,j}^x = J_{i,j}^y = -\cfrac12 J_{i,j}^z & = \cfrac{\gamma_F^2 (1-3\cos{^2\theta_{ij}})}{r_{ij}^3}, \label{c1}\\ 
 \tilde{\tilde{{J}}}_{k,l}^x = \tilde{\tilde{{J}}}_{k,l}^y = -\cfrac12 \tilde{\tilde{{J}}}_{i,j}^z & = \cfrac{\gamma_P^2 (1-3\cos{^2\theta_{kl}})}{r_{kl}^3}, \label{c2}\\
 \tilde{J}_{i,j}^x = \tilde{J}_{i,j}^y & = 0, \label{c3}\\
 \tilde{J}_{i,j}^z & = \cfrac{\gamma_F\gamma_P (1-3\cos{^2\theta_{ik}})}{r_{ik}^3}.\label{c4}
\end{align}

As in the homonuclear case, the $^{19}\text{F}$ FID is proportional to $C_x(t)$ defined by Eq.~\eqref{corf1}, but the dynamics is now determined by the Hamiltonian~\eqref{truncated_dipolar_unlike}.

\subsection{Lattice disorder}

The main type of defects in fluorapatite is the substitutions of $\chem{F}^-$ ions by other $\chem{X}^-$ ions.
Usually, $\chem{F}^-$ ions are substituted by $\chem{Cl}^-$ ions or by hydroxyl groups $(\chem{OH})^-$ \cite{Leroy-01}.
The presence of defects disrupts the fluorine chains and, in principle, leads to an adjustment of the positions of the neighbouring atoms.
In addition, both the stable isotopes of chlorine and protons of the $(\chem{OH})^-$ group are magnetically active. While the gyromagnetic ratio of chlorin nuclei is relatively small and hence the respected site can be treated as non-magnetic vacancy, 
the strongly magnetic proton spins would generate the inhomogeneous broadening of the $z$-components of the local magnetic fields sensed by the neighboring $^{19}\chem{F}$ and $^{31}\chem{P}$ nuclei.
In principle, an accurate calculation of the $^{19}\chem{F}$  FID should account for all of such effects. However, to the best of our knowledge, there is no detailed data about the concentrations of different types of defects in the sample used in the experiment \cite{Engelsberg-73}. Therefore, we choose to follow the approach of Ref.\cite{Cho-96}, namely, we assume that the fluorine atoms in fluorapatite are randomly replaced by non-magnetic substitutions with probability $\rho$.  The concentration of non-magnetic substitutions is then determined by matching the experimentally measured second moment of the FID. 

Let us consider some particular realization of disorder in the system. It can be specified by introducing a set of independent random binary variables $\{p_i\}$, which take value $0$ with probability $\rho$ and value $1$ with probability $(1-\rho)$. Here $i$ is the index of the fluorine lattice site and $\rho$ is the concentration of defects.
The values $p_i=1$ or $p_i=0$ correspond to the spin being either present or absent on site $i$ respectively.
As a result, in the full truncated dipolar-dipolar Hamiltonian given by Eq.~\eqref{truncated_dipolar_unlike}, spin operators $S_i^\alpha$ are substituted by $p_iS_i^\alpha$.

The definition of the auto-correlation function measured in experiment (\ref{corf1},\ref{MM}) is now changed to
\begin{equation}
 C_x(t)\equiv {\left\langle\cfrac{1}{D^{\{p\}}}\ttr{e^{i\BH t}M_x^{\{p\}}e^{-i\BH t} M_x^{\{p\}}}\right\rangle_{\{p\}}},
 \label{Cxd}
\end{equation}
where $\{p\}$ denotes a particular realization of disorder, $\langle\dotsb\rangle_{\{p\}}$ is the disorder average, and
\begin{equation}
 M_x^{\{p\}} = \lsum_i p_i S_i^x .
\end{equation}

According to the hybrid scheme,  the correlation function (\ref{Cxd}) is then reexpressed as a counterpart of Eq.(\ref{hybrid_autocorr}):
 \begin{equation} 
  C_{x} (t) = \cfrac{N_\BL}{N_{\BQ^\prime}}\left[ \BM^{\{p\}}_x(t) \BM'^{\{p\}}_x(0) \right]_{i.c., \{p\} },
  \label{hybrid_autocorr_dis}
\end{equation}
where
\begin{equation}
  \BM^{\{p\}}_x(t) = \sqrt{D^{\{p\}}_\BQ+1} \ \langle\psi(t)|\lsum_{i\in\BQ} p_i {S}_i^x |\psi(t)\rangle
  + \lsum_{m\in\BC} p_m s_m^x(t),
  \label{total_spin_dis}
  \end{equation}
  and 
\begin{equation}
  \BM^{\prime\{p\}}_x(t) = \sqrt{D^{\{p\}}_\BQ+1} \ \langle\psi(t)|\lsum_{i\in\BQ'} p_i {S}_i^x |\psi(t)\rangle .
  \label{subtotal_spin_dis}
  \end{equation}

\subsection{Simulations vs. experiment for fluorapatite}\label{fluor_sim}

\begin{figure}[t]
\begin{center}
 \includegraphics[width = 3.4in]{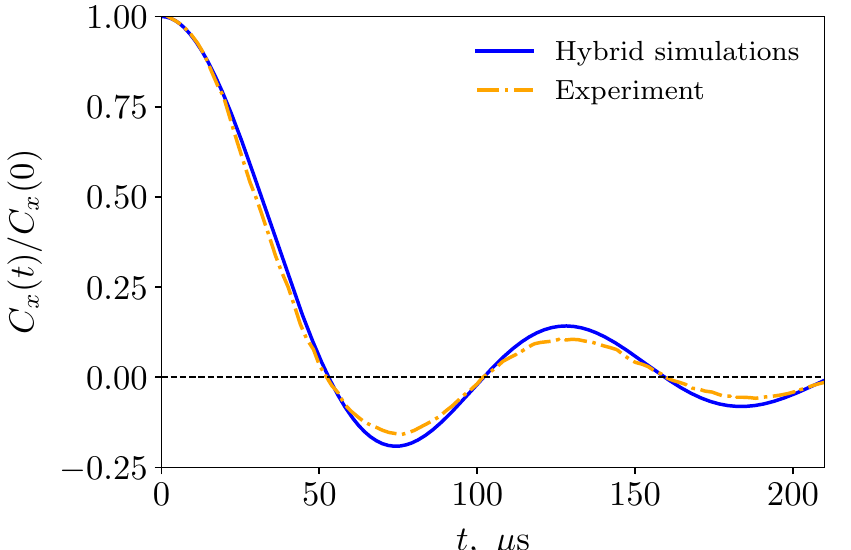}
 \caption{$^{19}\chem{F}$ FID in fluorapatite. Comparison of the results of the hybrid simulations including fluorine spins,  phosphorous spins and lattice disorder with the experimental data of Engelsberg {\it et al.}\cite{Engelsberg-73}.}\label{fluor_fid}
 \end{center}
\end{figure}

The comparison of the results of hybrid simulations with the experimental data of Ref.\cite{Engelsberg-73} is presented in Fig.~\ref{fluor_fid}. We used the concentration of fluorine vacancies  $\rho=0.077$ obtained by fitting of the experimental  second  moment of the FID with the theoretical value computed from first principles  (see the Supplementary Material \cite{supplement}). 
The size of the simulated hybrid lattice  was $9\times9\times7$ basis cells. The central quantum cluster $\BQ$ was chosen as a single chain of fluorine spins extending along the $c$-axis: it covered 7 basis cells and, therefore, included 14 fluorine spins. The rest of the spins were simulated classically. Since the quantum cluster $\BQ$ in this case was closed periodically, all its spins were equivalent with respect to the classical environment. Hence, each of the 14 lattice sites belonging to $\BQ$ also belonged to the subset $\BQ^\prime$ appearing in the definition (\ref{hybrid_autocorr_dis}) of the hybrid correlation function. 
We also performed hybrid simulations for a smaller system of $9\times9\times6$ basis cells with the central quantum cluster of 12 fluorine spins. The difference between the computed FIDs for the two hybrid lattices is smaller than the thickness of the plot lines in Fig.~\ref{fluor_fid}.  This implies that the above result amounts to a quantitatively reliable prediction of the FID of the fully quantum dynamics for the given interaction Hamiltonian and the chosen model of the lattice disorder. 
In Appendix~\ref{extra_fluorapatite}, we also present the results of hybrid FID calculations without the lattice disorder for an isolated fluorine chain and  for a three-dimensional lattice  with and without phosphorus nuclear spins.



Overall, the agreement between the numerical and the experimental results shown in Fig.~\ref{fluor_fid}  is good.
However, there is still a discrepancy, which, while being small in absolute terms, is larger than the predictive uncertainty estimate for the hybrid simulations. Therefore, this discrepancy is  due to either the experimental uncertainty or the approximate character of our lattice disorder model, which we had to resort to in the absence of more detailed information about the material.


\section{Conclusions}\label{conclusions}

Overall, we observe good quantitative agreement of the hybrid simulations of FIDs in $^\text{29}\text{Si}$-enriched silicon and in  fluorapatite with experiments. Both settings are characterized by low effective number of interacting neighbors $n_\text{eff}$ of each nuclear spin and, hence, are crucial for testing the predictive performance of the hybrid method as far as the essentially quantum aspects of the FID behavior are concerned. 
However, the conclusive assessement of the predictive power of the hybrid method is hindered by the experimental uncertainties and/or by the insufficient knowledge about lattice disorder, including vacancies, substitutions and paramagnetic impurities. More accurate NMR experiments on better characterized samples with small $n_\text{eff}$  need to be performed in order to conduct more stringent tests of the hybrid method. We also introduced the coupled quantum clusters method, which was shown to exhibit excellent performance when applied to the FID in $^\text{29}\text{Si}$-enriched silicon with the orientation of the external magnetic field imposing the natural partition of the full lattice into pairs of strongly coupled spins. Given these promising results, the performance of the latter method for a broader class of systems merits further systematic investigation.


\acknowledgments

This work was supported by a grant of the Russian Science Foundation (Project No. 17-12-01587).

\appendix

\section{Additional calculations and discussion of FIDs in $^{29}\text{Si}$-enriched silicon}
\label{extra_si_sims}

\begin{figure*}[t]
 \begin{center}
\includegraphics[width = 6.8in]{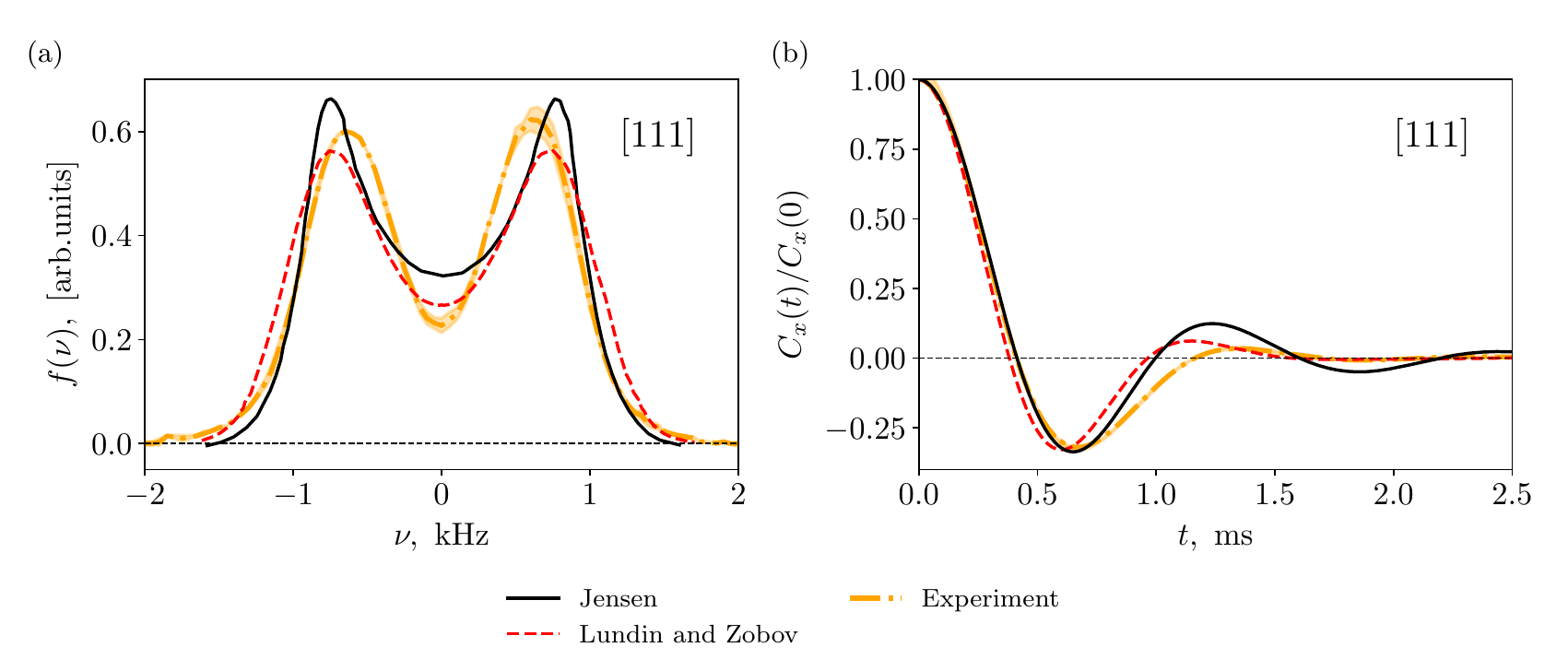}
  \caption{(a) Absorption lineshape and (b) FID  in $^\text{29}$Si diamond for $\mathbf{B}_0$ along $[111]$ crystal direction:
  comparison of the theoretical predictions of Jensen\cite{Jensen-95} and of Lundin and Zobov\cite{Lundin-18} with the experiment of Verhulst {\it et al.}\cite{Verhulst-03}. The experimental lines are the same as in Fig.~\ref{diamond_111}. 
  }
  \label{diamond_111_theory}
 \end{center}
\end{figure*}

In order to facilitate the comparison between the performance of the hybrid method for the calculation of $^{29}\text{Si}$ FID with $\mathbf{B}_0 || [111]$  as presented in Fig.~\ref{diamond_111} with the performance of alternative theoretical methods, we include in Fig.~\ref{diamond_111_theory} the same kind of theory-vs.-experiment plots for the (properly rescaled) theoretical predictions of Jensen \cite{Jensen-95} and of Lundin and Zobov \cite{Lundin-18}.


We also computed $^{29}\text{Si}$ FIDs for the external magnetic field $\mathbf{B}_0$ parallel to $[011]$, $[001]$.
The values of $n_\text{eff}$ in these two cases are $5.9$ and $27.4$ respectively.
According to the investigations of Elsayed and Fine~\cite{Elsayed-15}, the classical simulations are expected to perform well when $n_\text{eff}>4$.
Thus, following the argumentation of Ref.\cite{Starkov-18}, we estimate the uncertainty of the hybrid simulations by comparing their results with the results of purely classical simulations.
The schemes of the quantum clusters used in the hybrid simulations are shown in Fig.~\ref{cluster_schemes_extra}.
For the details of the simulations, see the Supplementary Material \cite{supplement}.

\begin{figure*}[t]\setlength{\unitlength}{0.1in}
\begin{center}
 \begin{picture}(60,20)
  \put(-2,0){\includegraphics[width=3.5in]{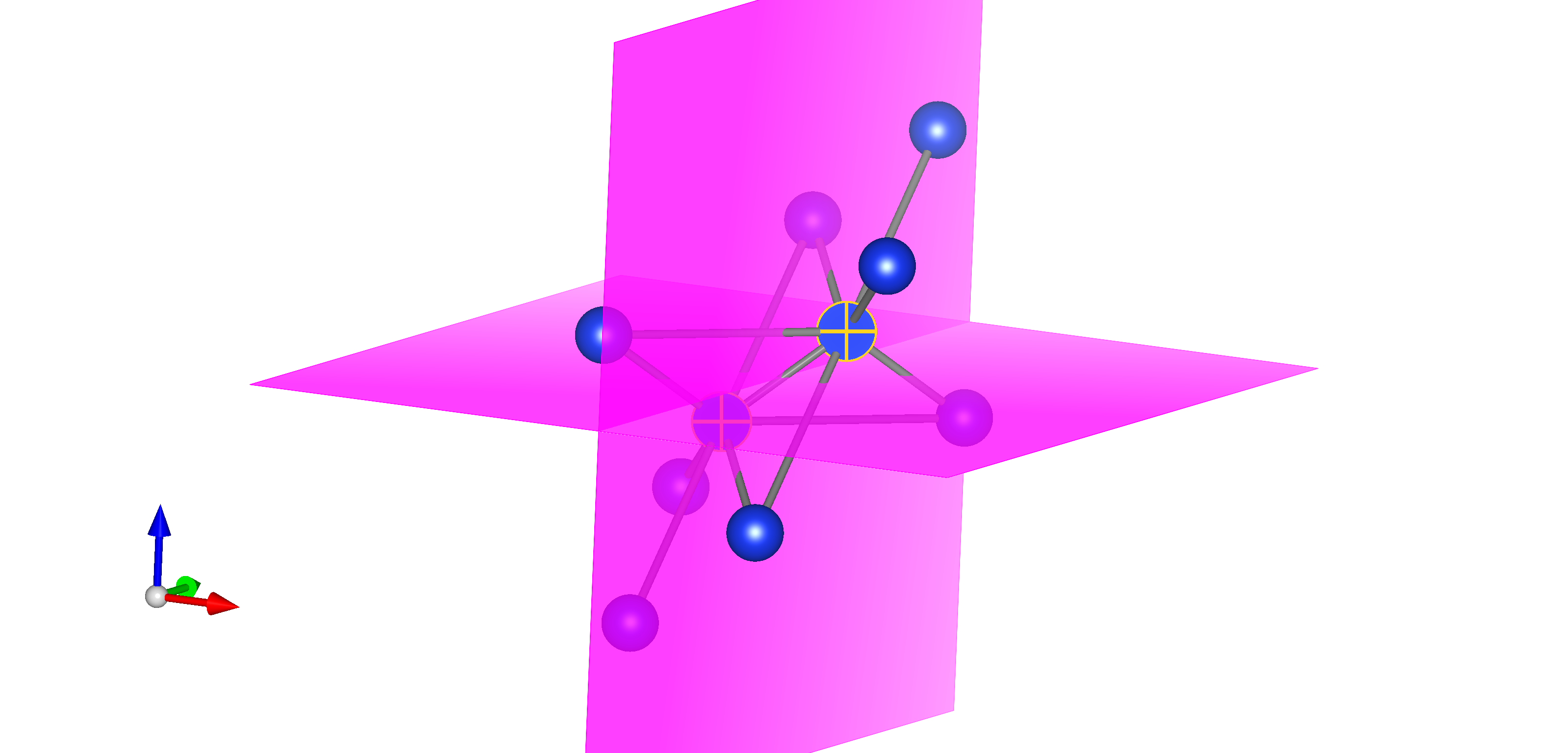}}
  \put(28,0){\includegraphics[width=3.5in]{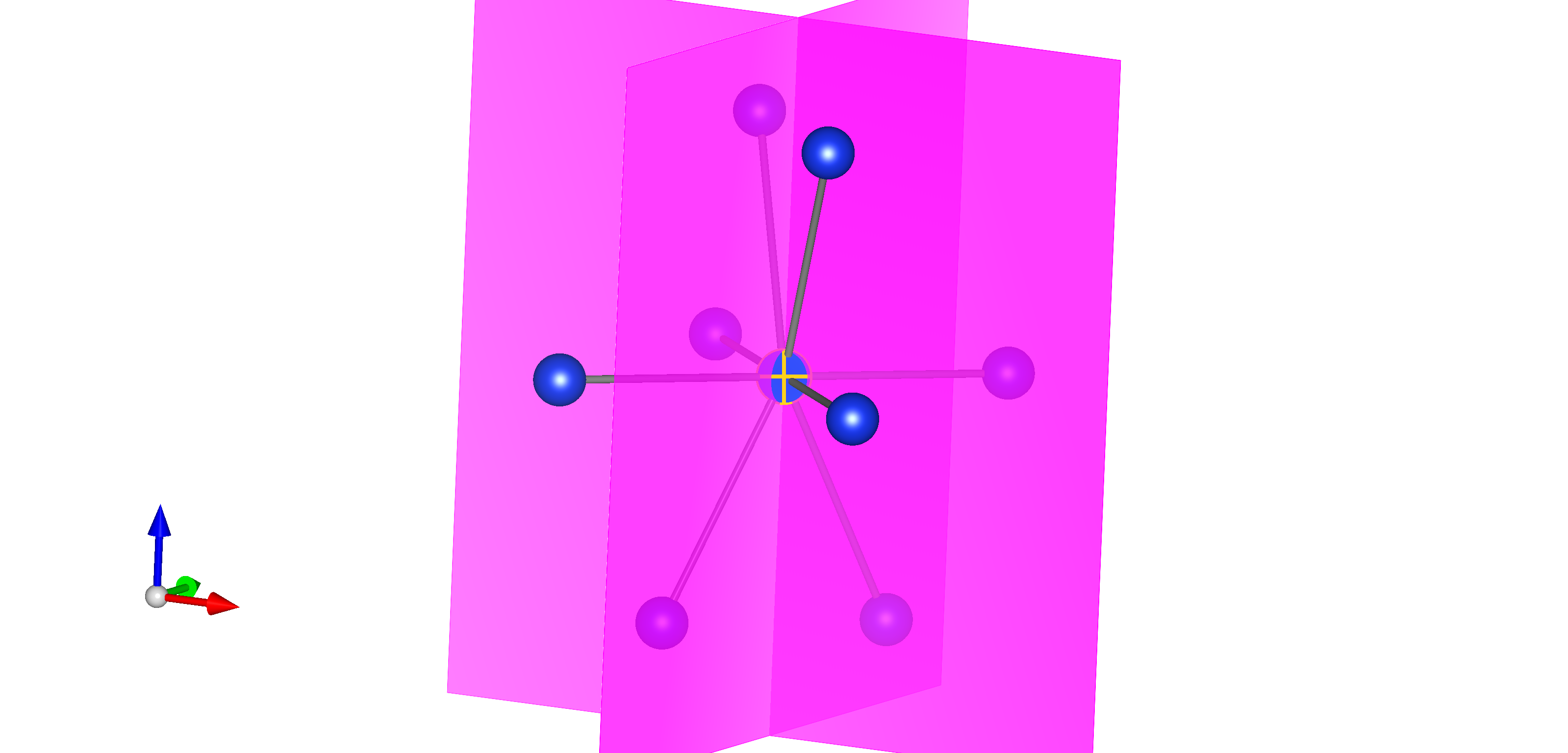}}
  \put(1,16){\pmark{a}}
  \put(31,16){\pmark{b}}
  
  \put(3.6,2.9){$\hat{a}$}
  \put(2.5,3.9){$\hat{b}$}
  \put(1.3, 5.8){$\hat{c}$}
  
  \put(33.6,2.9){$\hat{a}$}
  \put(32.5,3.9){$\hat{b}$}
  \put(31.3, 5.8){$\hat{c}$}

 \end{picture}
 \caption{Schemes of the quantum clusters for the hybrid simulations of $^{\text{29}}\text{Si}$-enriched silicon presented in Figs.~\ref{diamond_011} and \ref{diamond_001}:
 (a) for $\mathbf{B}_0\parallel [011]$;  (b) for $\mathbf{B}_0\parallel [001]$. Sites belonging to the subset $\BQ'$ are marked with $+$: two in (a) and one in (b).
 }\label{cluster_schemes_extra}
\end{center}
\end{figure*}
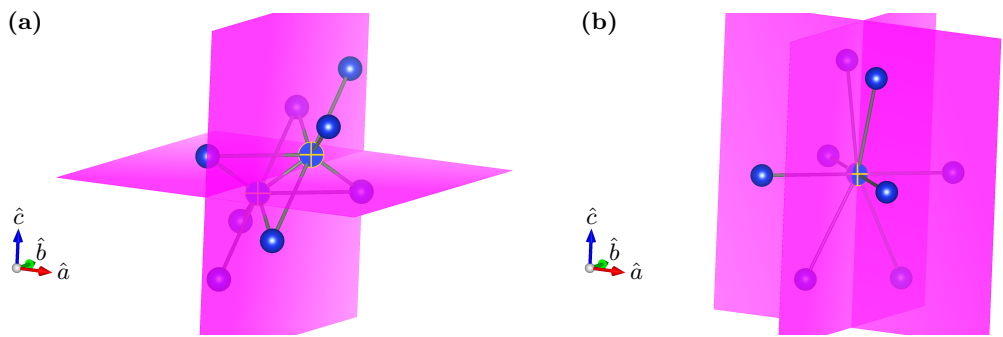

\begin{figure*}[t]
 \begin{center}
  \includegraphics[width = 6.8in]{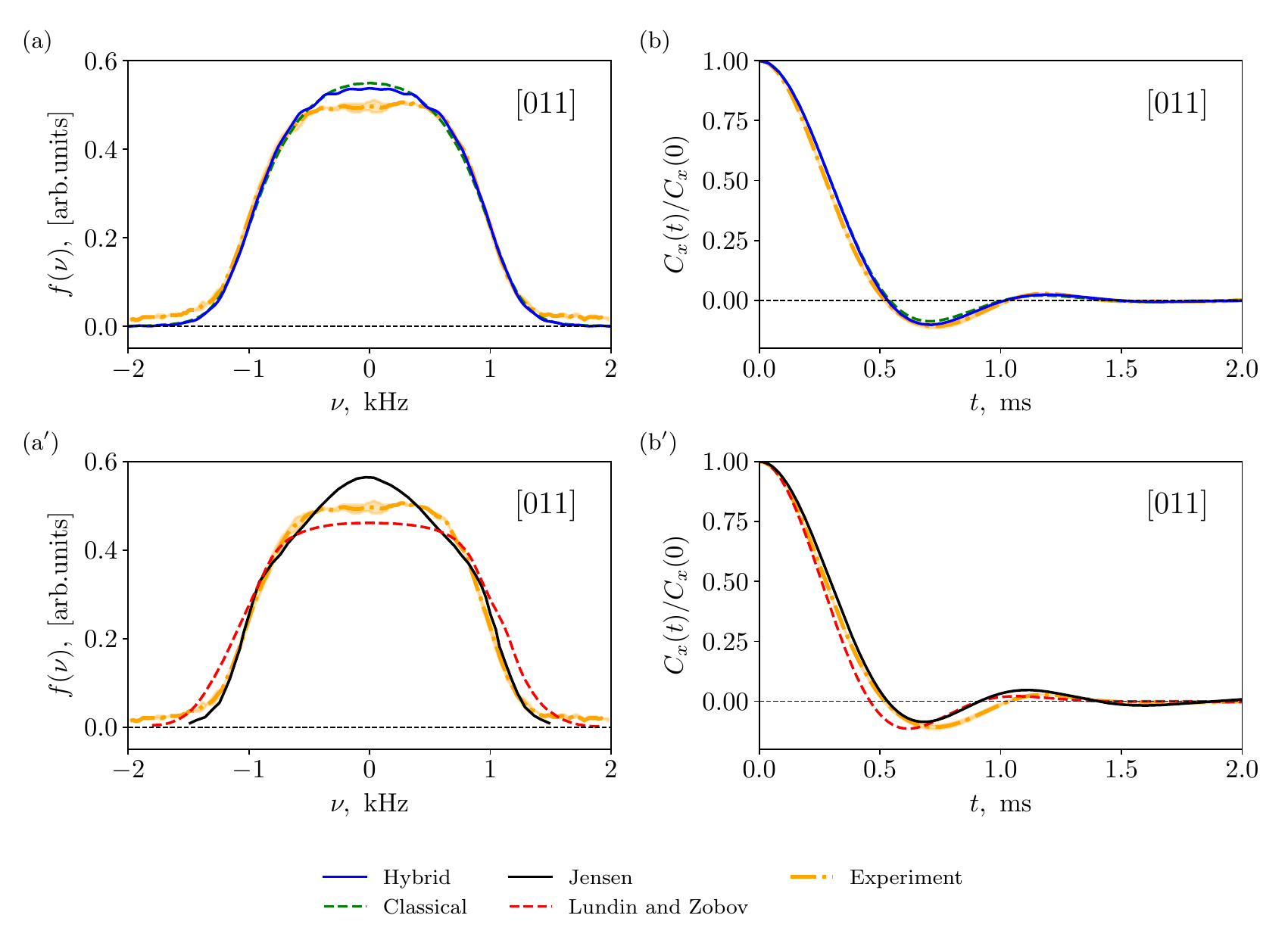}
  \caption{(a, a$'$) Absorption lineshapes  and (b, b$'$) FIDs  in $^\text{29}$Si-enriched silicon for $\mathbf{B}_0 \parallel [011]$.
  (a, b): comparison of the results of hybrid simulations with the experiment of Verhulst {\it et al.}\cite{Verhulst-03}
  (a$'$, b$'$): comparison of the theoretical predictions of Jensen\cite{Jensen-95} and of Lundin and Zobov\cite{Lundin-18} with the same experimental data.
  The scheme of the quantum cluster used in the hybrid simulations is displayed in Fig.~\ref{cluster_schemes_extra}(a).}\label{diamond_011}
 \end{center}
\end{figure*}

\begin{figure*}[t]
 \begin{center}
  \includegraphics[width = 6.8in]{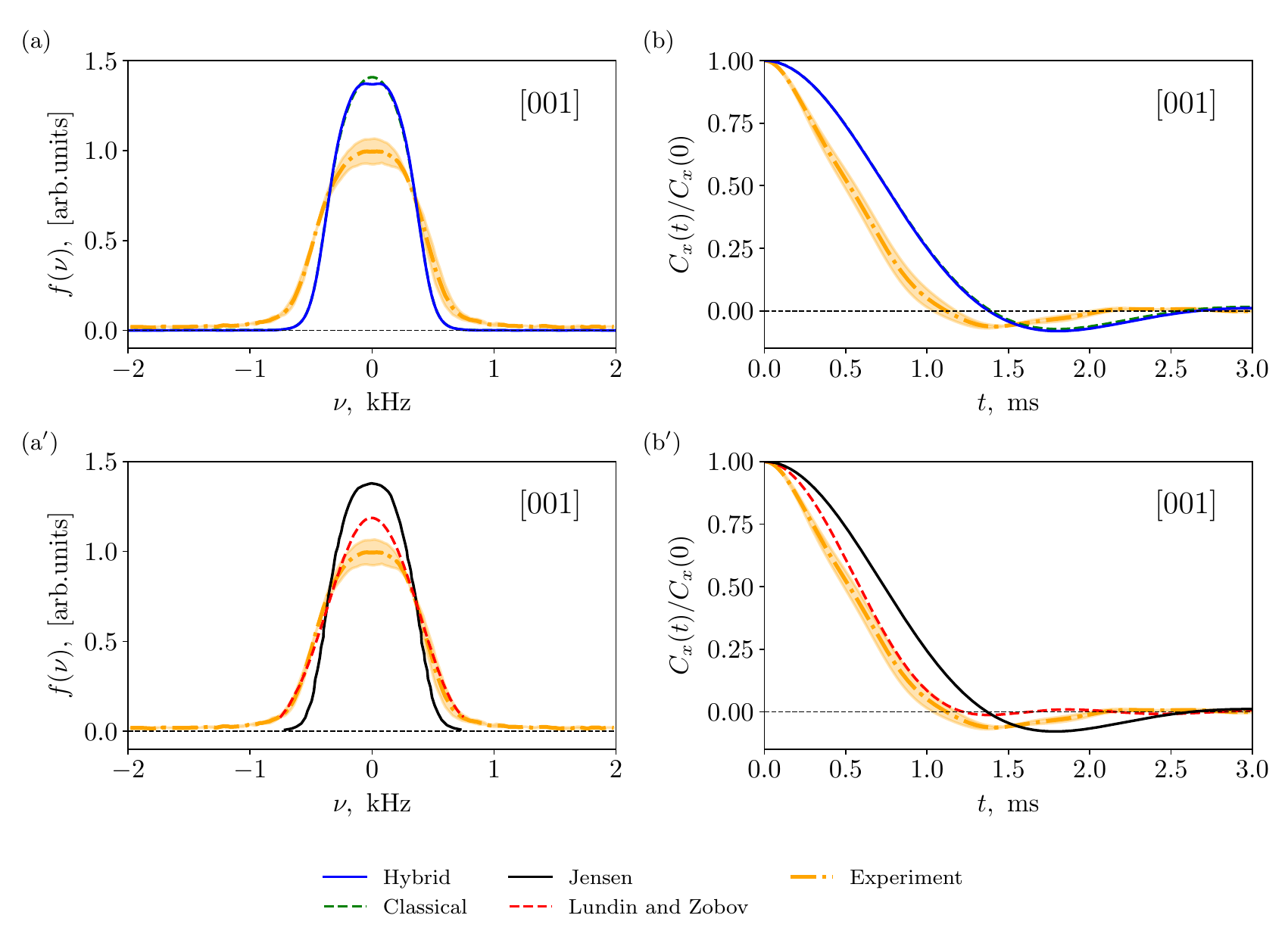}
  \caption{(a, a$'$) Absorption lineshapes  and (b, b$'$) FIDs  in $^\text{29}$Si-enriched silicon for $\mathbf{B}_0 \parallel [001]$.
  (a, b): comparison of the results of hybrid simulations with the experiment of Verhulst {\it et al.}\cite{Verhulst-03}
  (a$'$, b$'$): comparison of the theoretical predictions of Jensen\cite{Jensen-95} and of Lundin and Zobov\cite{Lundin-18} with the same experimental data.
  The scheme of the quantum cluster used in the hybrid simulations is displayed in Fig.~\ref{cluster_schemes_extra}(b).}\label{diamond_001}
 \end{center}
\end{figure*}

The results of our simulations are presented in Figs.~\ref{diamond_011} and~\ref{diamond_001} for external magnetic field $\mathbf{B}_0$ along $[011]$ and $[001]$ crystal directions respectively. The figures have identical structure.
The upper row [frames (a) and (b)] presents the comparison of the results of our simulations with experimental data of Verhulst {\it et al.} \cite{Verhulst-03}.
The lower row [frames (a\tprime) and (b\tprime)] presents the comparison of the theoretical predictions of Jensen \cite{Jensen-95} and of Lundin and Zobov \cite{Lundin-18} with the same experimental data.

In the case of $\mathbf{B}_0 || [001]$, we are reasonably confident that  large the discrepancy between the hybrid predictions and the experiments is due to the experimental uncertainties, which is evidenced by the fact that the ratio of the experimental and the theoretical second moments in this case is $2.73$. In the case of $\mathbf{B}_0 || [011]$, the second moment ratio is $1.4$, i.e. closer to 1, and hence the agreement between the hybrid calculations and the experiment is more satisfactory.

In the Supplementary Material, we also included the plots of the rescaled experimental results of Lefmann {\it et al.}\cite{Lefmann-94} for $^{13}$C-enriched diamond together with the data presented in Figs.~\ref{diamond_111}, \ref{diamond_111_theory}, \ref{diamond_011}, and \ref{diamond_001}.

\section{Additional calculations and discussion of FID in fluorapatite}
\label{extra_fluorapatite}

\begin{figure}[t]
\begin{center}
 \includegraphics[width = 3.4in]{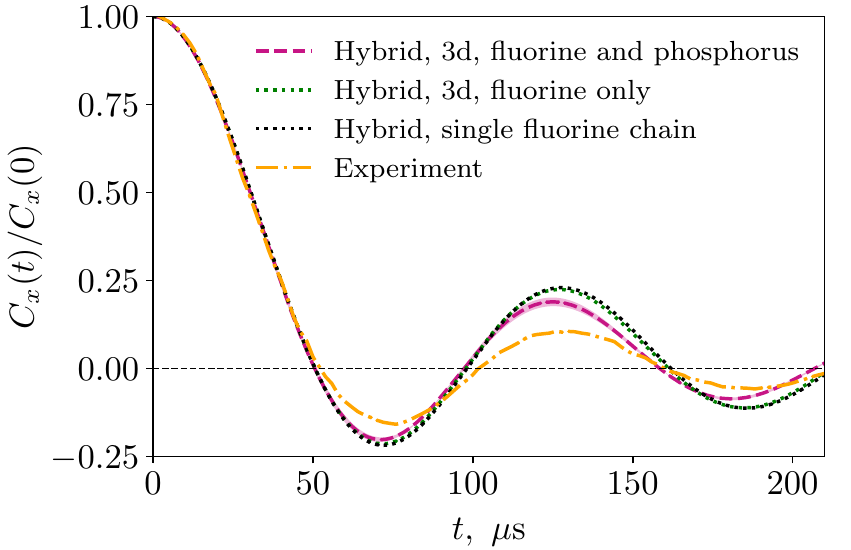}
 \caption{$^{19}\chem{F}$ FID in fluorapatite. Comparison of hybrid simulations with the experiment of Engelsberg {\it et al.}\cite{Engelsberg-73} for different levels of modeling defined in the text: black dotted line ---  single fluorine chain, green dotted line ---  three-dimensional lattice of fluorine nuclei only; magenta dashed line ---  three dimensional lattice of fluorine and phosphorous nuclei. None of the hybrid simulations presented in this figure included lattice disorder.
 }\label{fluor_fid2}
 \end{center}
\end{figure}

In Fig.~\ref{fluor_fid2}, we present the results of the hybrid simulations for different levels of modeling  $^{19}\chem{F}$ FID in fluorapatite. We consider a series of models, which  gradually become more realistic:
\begin{enumerate}
\renewcommand{\labelenumi}{(\roman{enumi})}
 \item isolated fluorine chain without disorder;
 \label{noud3}
 \item three-dimensional fluorine lattice without disorder and without phosphorus nuclear spins;\label{noud}
 \item three-dimensional lattice with phosphorus nuclei, but without disorder;\label{nod}
 \item three-dimensional lattice with phosphorus nuclei and disorder.\label{ud}
\end{enumerate}
Model (iv) corresponds to the simulations described in Section~\ref{fluor_section} and presented in Fig.~\ref{fluor_fid}.

The parameters of the simulated hybrid lattices are the following.
For model (i) the hybrid lattice was a chain of length $201$ with periodic boundary conditions; the size of the quantum cluster $\BQ$ was $12$ and the subset $\BQ'$ included two central spins of cluster $\BQ$.
For model (ii), the fluorine sublattice of the size $7\times7\times13$ spins was used, and both   $\BQ$ and  $\BQ'$ were chosen in the form of a periodic 13-spin fluorine chain.
The parameters for model (iii)  were the same as for the model (iv) used in Section~\ref{silicon_diamond}: the lattice size was $9\times9\times7$ basis cells, while $\BQ$ and  $\BQ'$  were chosen as a  14-spin fluorine chain.

We, finally, note that the results of the simulations for models (i) and (ii) nearly coincide with each other, thereby corroborating the quasi-one-dimensional character of the fluorine sublattice.




\FloatBarrier

\bibliography{ClassSimuNew}
\onecolumngrid
\clearpage

\begin{center}
  \bf\large
  Supplementary Material\\ for the article\\ ``Free induction decays in nuclear spin-1/2 lattices with small number of interacting neighbors: the cases of silicon and fluorapatite''
\end{center}\bigskip
  
\twocolumngrid
  
\setcounter{figure}{0}
\setcounter{equation}{0}
\setcounter{table}{0}
\setcounter{section}{0}
  
\renewcommand{\thefigure}{S\arabic{figure}}

\renewcommand{\theequation}{S\arabic{equation}}

\renewcommand{\thetable}{S\arabic{table}}

\renewcommand{\thesection}{S\Roman{section}}

All equation numbers, figure numbers and reference numbers without prefix ``S'' refer to the respective numbers in the main article.

\section{Details of the simulations for $^{29}\text{Si}$-enriched silicon}

For the orientation of external magnetic field $\mathbf{B}_0$ along the $[111]$ crystal direction, we specified the lattice with the help of the original two-site basis cell and the original set of primitive vectors (see Eqs.~\ref{simple_vectors} and~\ref{simple_basis}). The full lattice size for the hybrid and the classical simulations was $9\times9\times9$ two-site cells, which included $1458$ lattice sites in total.
In this particular case, the strongest interaction is between the spins of the same two-site cell.
The next in strength are the interactions between the nearest spins of the different two-site cells.
The schemes of cluster 1 and cluster 2 are presented in Fig.~\ref{cluster_schemes_111} (a) and (b) respectively.
We chose the quantum cluster 1 to contain a two-site cell and its translations by the primitive vectors and their inverses, 7 pairs in total.
In the hybrid FID calculation, the central spin pair of the above cluster formed subset $\BQ'$ entering Eq.\eqref{central_spins}.
Cluster 2 was obtained from cluster 1 by leaving only three two-node cells obtained by primitive vectors translations; all three pairs belonged to the subset $\BQ'$.

For $\mathbf{B}_0$ along $[001]$ and $[011]$ crystal directions, we used the freedom to choose the basis and the primitive vectors to specify the full lattice. Instead of the two-site basis cell given by Eq.~\eqref{simple_basis}, we used an eight-site basis cell containing the original two-site basis cell and its three translations by the original primitive vectors  given by Eq.~\eqref{simple_vectors}:
\begin{equation}
\begin{array}{llll}
 \mathbf{v\mprime}_0 = \mathbf{v}_0,   &  \mathbf{v\mprime}_2 = \mathbf{v_0} + \mathbf{l}_1,  &  \mathbf{v\mprime}_4 = \mathbf{v_0} + \mathbf{l}_2,  &  \mathbf{v\mprime}_6 = \mathbf{v_0} + \mathbf{l}_3,  \\
 \mathbf{v\mprime}_1 = \mathbf{v}_1,   &  \mathbf{v\mprime}_3 = \mathbf{v_1} + \mathbf{l}_1,  &  \mathbf{v\mprime}_5 = \mathbf{v_1} + \mathbf{l}_2,  &  \mathbf{v\mprime}_7 = \mathbf{v_1} + \mathbf{l}_3.
\end{array}\label{new_basis}
\end{equation}
The corresponding set of primitive vectors was
\begin{equation}
 \mathbf{l\mprime}_1 = a_0\hat{a}, \quad \mathbf{l\mprime}_2 = a_0\hat{b}, \quad \mathbf{l\mprime}_3 = a_0\hat{c}.
\end{equation}
The full lattice size for the hybrid and the classical simulations was $7\times7\times7$ eight-site basis cells, which included $2744$ lattice sites in total. Periodic boundary conditions were used with respect to the new set of primitive vectors.

For the orientation of external magnetic field $\mathbf{B}_0$ along the $[011]$ crystal direction, we chose two  sites of the basis cell specified by $\mathbf{v\mprime}_0$ and $\mathbf{v\mprime}_1$ to form the subset $\BQ'$ in Eq.~\eqref{central_spins}.
The quantum cluster $\BQ$ was comprised of the above two sites and eight more sites where spins had the strongest coupling with the spins on the two ``primary'' sites.
With the origin at the site specified by $\mathbf{v\mprime}_0$, the coordinates of the spins of the quantum cluster $\BQ$ in units of $a_0$ are:
\begin{equation}
 \begin{array}{ll}
 [0.0, 0.0, 0.0],   &   [0.25, 0.25, 0.25], \\ \relax
 [0.5, 0.0, 0.5],   &   [0.0, 0.5, 0.5],    \\ \relax
 [0.5, 0.5, 0.0],   &   [0.25, 0.75, 0.75], \\ \relax
 [-0.25, 0.25, -0.25],  &   [-0.25, -0.25, 0.25], \\ \relax
 [0.25, -0.25, -0.25],  &   [0.0, -0.5, -0.5].
 \end{array}
\end{equation}
The scheme of the cluster is presented in Fig.~\ref{cluster_schemes_extra} (a).

For the orientation of external magnetic field $\mathbf{B}_0$ along the $[001]$ crystal direction, we chose only one site of the basis cell specified by $\mathbf{v\mprime}_0$ to form the subset $\BQ'$ in Eq.~\eqref{central_spins}. In order to construct the quantum cluster $\BQ$, we took the above site
and added 8 sites, where the spins had the strongest interaction with the one on the ``primary'' site.
With the origin of the coordinate system at the central site, the coordinates of the sites of cluster $\BQ$ in units of $a_0$ are:
\begin{equation}
 \begin{array}{ll}
[0.0, 0.0, 0.0],     &   [-0.5, -0.5, 0.0],\\ \relax
[0.5, -0.5, 0.0],    &   [-0.5, 0.5, 0.0], \\ \relax
[0.5, 0.5, 0.0],     &   [-0.25, -0.25, -0.75], \\ \relax
[0.25, -0.25, 0.75], &   [-0.25, 0.25, 0.75], \\ \relax
[0.25, 0.25, -0.75]. & \relax
\end{array}\label{xxx}
\end{equation}{}
The scheme of the cluster is presented in Fig.~\ref{cluster_schemes_extra}~(b).

\section{Procedure for generating experimental plots for $^\text{29}\text{Si}$-enriched silicon}

 Verhulst {\it et al.}\cite{Verhulst-03} reported several measurements for external magnetic field oriented along crystal directions $[111]$, $[011]$ and $[001]$ and along their symmetric equivalents.
The measurements along pairs of equivalent directions still produced slightly different results.
These differences might be, e.g., attributable to the uncertainties in the orientation of the sample.
In order to use both measurements for a pair of eqivalent directions, 
we  took the experimental curve for the indicated direction ($[111]$, $[011]$ or $[001]$) and a curve for an equivalent direction and then plotted their half-sum as the curve labeled ``experiment''.  The experimental uncertainty was then represented graphically as the area between the two experimental curves contributing to the average. Such a quantification of uncertainty does not include systematic experimental errors.

\section{Comparison of the results for $^\text{29}\text{Si }$ FIDs  with the experimental $^\text{13}\text{C}$ FIDs in diamond measured by Lefmann {\it et al.} }

\begin{figure*}[t]
 \begin{center}
  \includegraphics[width = 6.8in]{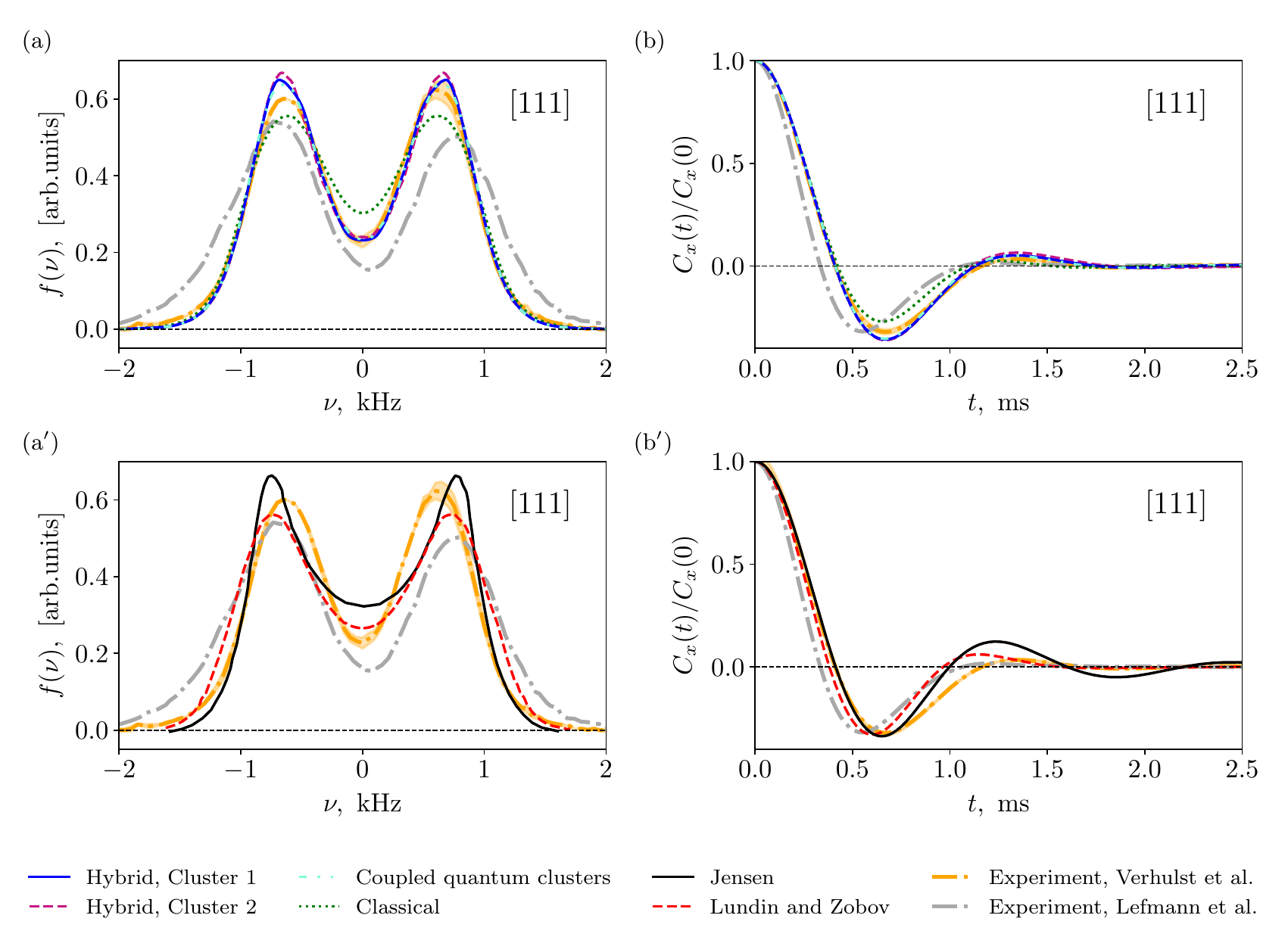}
  \caption{(a, a$'$) Absorption lineshapes  and (b, b$'$) FIDs  in $^\text{29}$Si-enriched silicon for $\mathbf{B}_0 \parallel [111]$. The content of these plots is the same as that of Figs.~\ref{diamond_111} and \ref{diamond_111_theory} of the main article with the only difference being that the lines representing the rescaled experimental data of  Lefmann {\it et al.}\cite{Lefmann-94} for the absorbtion lineshape and the corresponding FID in $^\text{13}$C-enriched diamond  were added.}\label{diamond_111_extra}
 \end{center}
\end{figure*}

\begin{figure*}[t]
 \begin{center}
  \includegraphics[width = 6.8in]{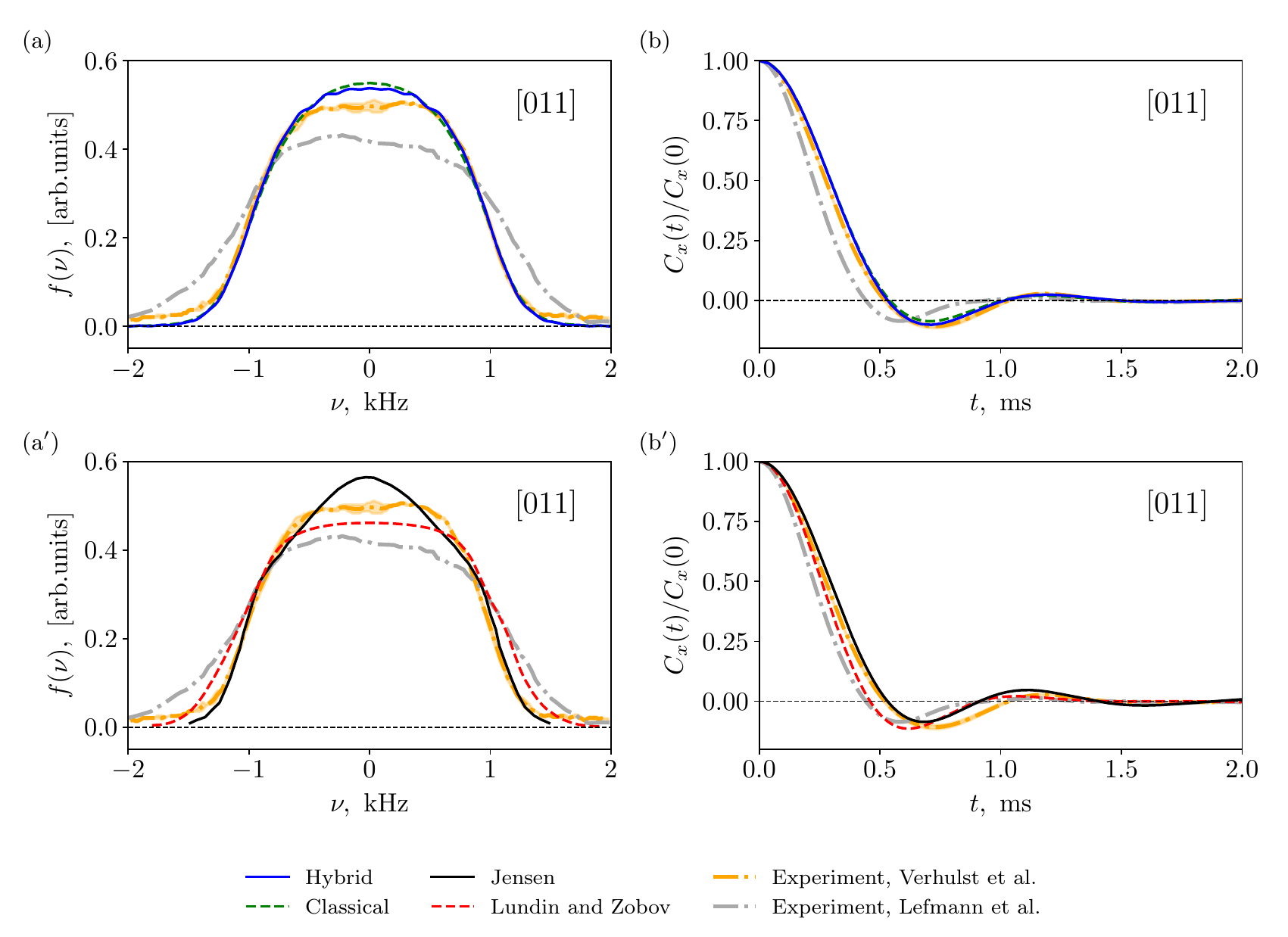}
  \caption{(a, a$'$) Absorption lineshapes  and (b, b$'$) FIDs  in $^\text{29}$Si-enriched silicon for $\mathbf{B}_0 \parallel [011]$. The content of these plots is the same as that of Fig.~\ref{diamond_011} of the main article with the only difference being that the lines representing the rescaled experimental data of  Lefmann {\it et al.}\cite{Lefmann-94} for the absorbtion lineshape and the corresponding FID in $^\text{13}$C-enriched diamond  were added.}\label{diamond_011_extra}
 \end{center}
\end{figure*}

\begin{figure*}[t]
 \begin{center}
  \includegraphics[width = 6.8in]{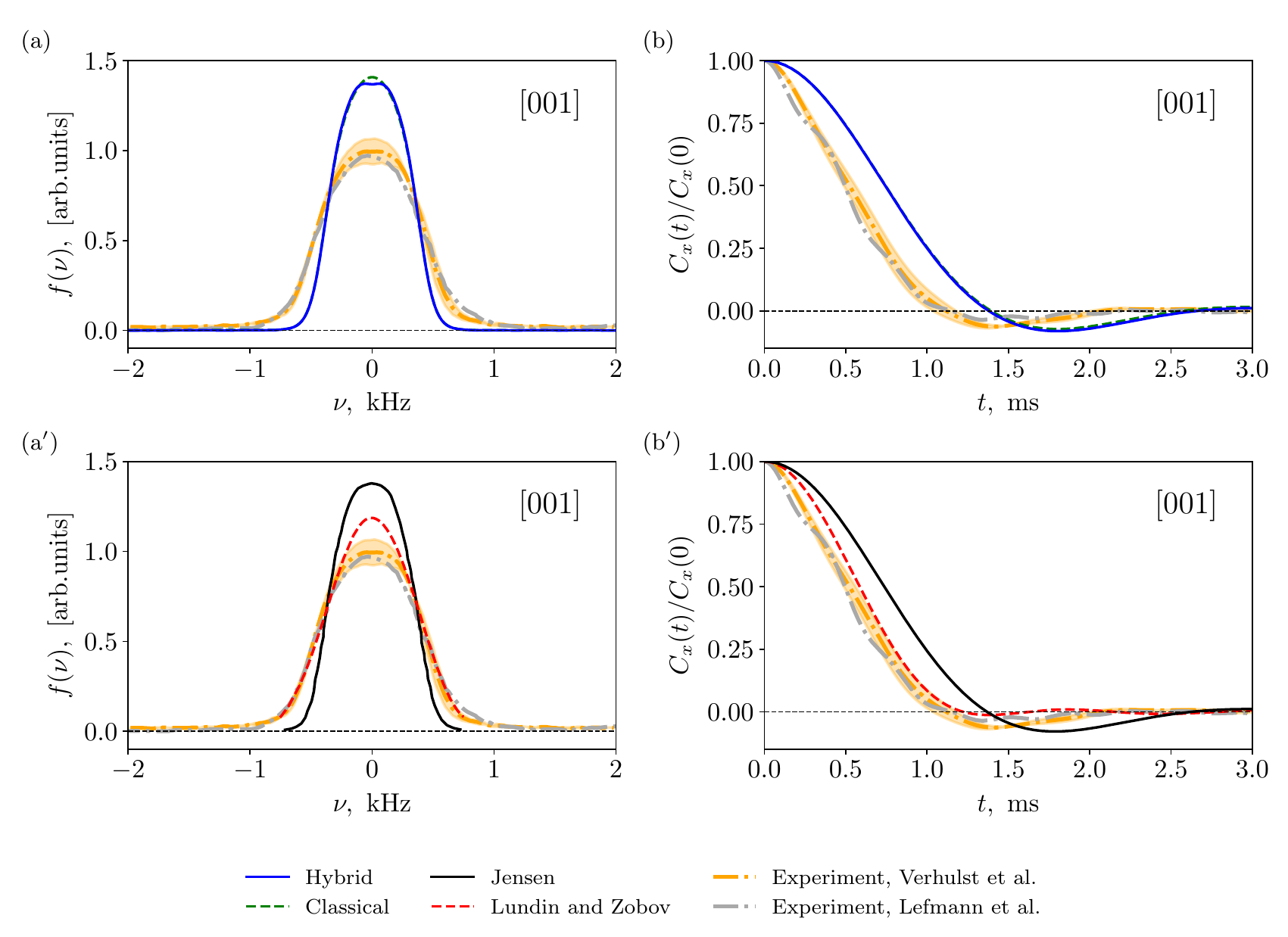}
  \caption{(a, a$'$) Absorption lineshapes  and (b, b$'$) FIDs  in $^\text{29}$Si-enriched silicon for $\mathbf{B}_0 \parallel [001]$. The content of these plots is the same as that of Fig.~\ref{diamond_001} of the main article with the only difference being that the lines representing the rescaled experimental data of  Lefmann {\it et al.}\cite{Lefmann-94} for the absorbtion lineshape and the corresponding FID in $^\text{13}$C-enriched diamond  were added.}\label{diamond_001_extra}
 \end{center}
\end{figure*}

In Figs. \ref{diamond_111_extra},~\ref{diamond_011_extra} and~\ref{diamond_001_extra}, we re-display the plots from Figs.~\ref{diamond_111}, \ref{diamond_111_theory},  \ref{diamond_011} and~\ref{diamond_001}  of the main article with the only difference that the experimental data of Lefmann {\it et al.}~\cite{Lefmann-94} for $^\text{13}\text{C}$ FIDs in  99\% $^\text{13}\text{C}$-enriched diamond are added. The time axis of the $^\text{13}\text{C}$ data is multiplied by the ratio ${ \gamma_{\text{C}}^2 a_{\text{\text{Si}}}^3 \over \gamma_{Si}^2 a_{\text{C}}^3 } $  to match the FID time scale of $^{31}\text{Si}$-enriched silicon. Here $\gamma_{\text{Si}}$, $\gamma_{\text{C}}$ are the respective gyromagnetic ratios, and  $a_{\text{Si}}$, $a_{\text{C}}$ the lattice periods.

In Table~\ref{m_diff}, we list the ratios  of the experimental  over the theoretical second moments $M_2^e /M_2^t$   for the data of Verhulst {\it et al.} and of Lefmann {\it et al.} for ${\bf B}_0$ along $[111]$, $[011]$ and $[001]$ crystal directions. The values of $M_2^t$ were computed from first principles for truncated magnetic dipolar interaction. 
For the $[111]$ and the $[011]$ directions, the second moments of the data of Verhulst {\it et al.} are closer to the theoretical values by roughly a factor of two.

In general, the deviations of the ratio $M_2^e / M_2^t$ from 1 seen in Table~\ref{m_diff}  can be either due to  experimental errors, including systematic ones, or due to the deviation of the actual microscopic setting from the perfectly periodic lattice with truncated magnetic dipolar interaction (\ref{dipolar}) between spins. In other words, this deviation quanitfies the uncertainty of the reference plots as opposed to the uncertainty due to the approximate character of the simulation method itself.

\begin{table}
 \begin{tabular}{|l|c|c|c|}\hline
   & $[111]$ & $[011]$ & $[001]$\\ \hline
$M_2^e /M_2^t$, Verhulst et.\ al. & $1.33$ & $1.41$ & $2.73$\\ \hline
$M_2^e /M_2^t$, Lefmann et.\ al. & $1.66$ & $1.72$ & $2.35$\\\hline
 \end{tabular}
\caption{Ratios $M_2^e /M_2^t$ of the experimental $M_2^e$ and the theoretical $M_2^t$ second moments for the data of Verhulst {\it et al.}cite{Verhulst-03} and Lefmann {\it et al.}cite{Lefmann-94} for three different orientations of ${\bf B}_0$.}\label{m_diff}
\end{table}

\begin{figure}
\begin{center}
 \includegraphics[width = 3.4in]{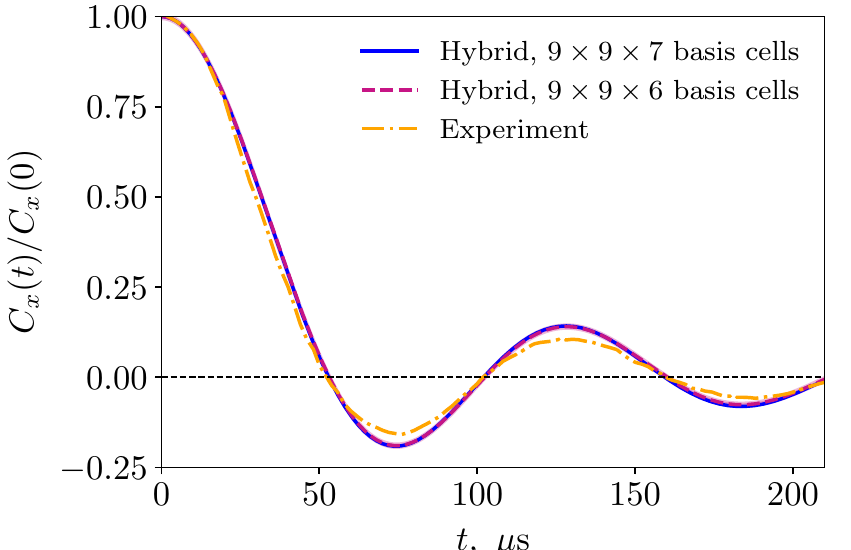}
 \caption{$^{19}\chem{F}$ FID in fluorapatite. Comparison of simulations for the hybrid lattices of different size. Blue line: hybrid lattice comprised of $9\times9\times7$ basis cells. Magenta dashed line: hybrid lattice comprised of $9\times9\times6$ basis cells. 
 }\label{fluor_fid3}
 \end{center}
\end{figure}

\section{Estimate of the concentration of fluorine vacancies {\large $\rho$} in fluorapatite}

In Section~\ref{fluor_sim} of the main article, we gave the estimate of the concentration of fluorine vacancies $\rho$ in fluorapatite obtained from the comparison of the theoretical and the experimental second moments of the FID. This was done as follows.

The second moment is defined as
\begin{equation}
 M_2 = -C^{\prime\prime}_x(0)/C(0),\label{moment2}
\end{equation}
where $C_x(t)$ is given by Eq.~\eqref{Cxd}.
Substituting the equation of motion for the operator $M_x^{\{p\}}(t)$ into \eqref{Cxd}, we obtain:
\begin{equation}
 M_2 = -\cfrac{\left\langle\ttr{\left[\BH, M_x^{\{p\}}\right]^2}\right\rangle_{\{p\}}}{\left\langle\left(M_x^{\{p\}}\right)^2\right\rangle_{\{p\}}},\label{moment2_exposed}
\end{equation}
where the Hamiltonian $\BH$ is given by Eq.~\eqref{truncated_dipolar_unlike} with $S_i^\alpha$ operators replaced by $p_i S_i^\alpha$.
Since $p_i$ are binary random variables,  $\langle p_i^n\rangle_{\{p\}} = (1-\rho)$ for any natural number $n$. This greatly simplifies the disorder averaging.

Evaluating Eq.(\ref{moment2_exposed}), we obtained:
\begin{equation}
 M_2 = (1-\rho)\cdot \cfrac{3}{4}S(S+1) \lsum_{j\neq i} (J_{i,j}^z)^2 + \cfrac{1}{3} I(I+1)\lsum_{k} (\tilde{J}_{i,k}^z)^2,
\end{equation}
where lattice index $i$ is fixed, while the summation is only over indices $j$ and $k$.
Here we took into account that $J_{i,j}^x = J_{i,j}^y = -\frac12 J_{i,j}^z$ and
$\tilde{J}_{i,k}^x = \tilde{J}_{i,k}^y = 0$.  
The only difference with the case where there is no disorder is that the contribution of fluorine nuclei is multiplied by the factor of $(1-\rho)$.
Since both $^{19}\text{F}$ and $^{31}\text{P}$ have spins 1/2,
we can finally write
\begin{equation}
 M_2 = (1-\rho) \ \cfrac{9}{16} \lsum_{j\neq i} (J_{i,j}^z)^2 + \cfrac{1}{4}\lsum_{k} (\tilde{J}_{i,k}^z)^2 = (1-\rho) M_{\text F} + M_{\text P},
 \label{M2MFMP}
\end{equation}
where $M_{\text F}$ and $M_{\text P}$ are defined through this equation.

Equation (\ref{M2MFMP}) allows us to express the disorder parameter $\rho$ as
\begin{equation}
 \rho = 1 - \cfrac{M_2-M_{\text P}}{M_{\text F}}.
 \label{rhoM}
\end{equation}
The experimentally measured second moment is now taken as $M_2$, while 
 $M_{\text P}$ and $M_{\text F}$ are computed numerically. This gives:
\begin{align}
 M_2 & = 1207.0\ \text{kHz}^2 \\
 M_{\text P} & = 34.6\ \text{kHz}^2 \\
 M_{\text F} & = 1270.6\ \text{kHz}^2
\end{align}
Equation (\ref{rhoM}) then yields $\rho =0.077$.

\section{Predictive uncertainty of the fluorapatite hybrid simulation results presented in Fig.~\ref{fluor_fid}}

In Fig.~\ref{fluor_fid3}, we compare the results of the hybrid simulations presented in Fig.~\ref{fluor_fid} of the main article with the simulations for the hybrid lattice of the size $9\times9\times6$ basis cells (including the phosphorus spins and disorder) with the central quantum cluster $\BQ$ comprised of 12 spins forming a periodic spin chain.
The hybrid lattice behind Fig.~\ref{fluor_fid3} was larger --- it consisted of $9\times9\times7$ basis cells with
quantum cluster $\BQ$ containing 14 spins.
As one can see in Fig.~\ref{fluor_fid3}, the deviation between the results of the simulations for the above two lattices is of the order of the thickness of the lines, which indicates high accuracy of the hybrid predictions for the underlying fully quantum lattice.

\

\section{Statistics behind the plots}

In Table \ref{runtable}, we list the number of computational runs behind the plots presented in both the main article and the supplementary material. Each run had length $10 \, T_0$, where $T_0$ is the maximum time, for which the correlation functions were eventually computed. The values of $T_0$ are also included in Table \ref{runtable}.

\begin{table*}[b]
 \begin{center}
 \small
  \begin{tabular}{|c|c|c|c|c|c|}\hline
   Material & ${\bf B}_0$ & Figure & Plot description & $T_0$ & Number of runs \\ \hline
    \multirow{8}{*}{$^{29}\text{Si}$}  & \multirow{4}{*}{$[111]$} & \multirow{4}{*}{\ref{diamond_111}(a,b), \ref{diamond_111_extra}(a,b)} & Hybrid, cluster 1 & 16.11 ms & 376903 \\ \cline{4-6}
                    &   &   & Hybrid, cluster 2 & 16.11 ms & 647527 \\ \cline{4-6}
                    &   &   & Coupled quantum clusters & 16.11 ms & 37601 \\ \cline{4-6}
                    &   &   & Classical & 16.11 ms & 348932 \\ \cline{2-6}
                    & \multirow{2}{*}{$[011]$} & \multirow{2}{*}{\ref{diamond_011}(a,b), \ref{diamond_011_extra}(a,b)} & Hybrid & 21.48 ms & 398896 \\ \cline{4-6}
                    &   &   & Classical & 10.74 ms & 168098 \\ \cline{2-6}
     & \multirow{2}{*}{$[001]$} & \multirow{2}{*}{\ref{diamond_001}(a,b), \ref{diamond_001_extra}(a,b)}& Hybrid & 21.48 ms & 382768 \\ \cline{4-6}
                    &   &   & Classical & 10.74 ms & 456384 \\ \hline
    \multirow{9}{*}{fluorapatite} & \multirow{9}{*}{$[001]$} & \ref{fluor_fid}, \ref{fluor_fid3} & \parbox{3.4cm}{\vrule width 0in height 3.5mm Hybrid, full 3D lattice\par\vspace{-0.5mm} with disorder,\par\vspace{-0.5mm} $9\times9\times7$ basis cells\vrule width 0in depth 1mm} & 210.5 $\mu$s & 281496 \\ \cline{3-6}
                    &   & \multirow{4}{*}{\ref{fluor_fid2}}  & \parbox{3.4cm}{\vrule width 0in height 3.5mm Hybrid, 3D lattice,\par\vspace{-0.5mm} fluorine and phosphorus \par\vspace{-0.5mm} without disorder\vrule width 0in depth 1mm} & 210.5 $\mu$s & 37868 \\ \cline{4-6}
                    &   &   & \parbox{3.4cm}{\vrule width 0in height 3.5mm Hybrid, 3D lattice,\par\vspace{-0.5mm} fluorine only\vrule width 0in depth 1mm} & 210.5 $\mu$s & 44878 \\ \cline{4-6}
                    &   &   & Hybrid, single fluorine chain & 210.5 $\mu$s & 93562 \\ \cline{3-6}
                    &   & \ref{fluor_fid3} & \parbox{3.4cm}{\vrule width 0in height 3.5mm Hybrid, full 3D lattice\par\vspace{-0.5mm} with disorder,\par\vspace{-0.5mm} $9\times9\times6$ basis cells \vrule width 0in depth 1mm} & 210.5 $\mu$s & 63116 \\ \hline
  \end{tabular}
\end{center}
\caption{The number of computational runs behind the plotted correlation functions. The time length of each run was $10 \, T_0$.}
\label{runtable}
\end{table*}

\end{document}